# Leveraging Time-Causal State Variable Aggregation for Real-Time Schedule of Massive Air Conditioners

Jingguan Liu, *Student Member, IEEE,* Xiaomeng Ai, *Member, IEEE,* Shichang Cui, *Member, IEEE,* Xizhen Xue, *Member, IEEE,* Shengshi Wang, *Student Member, IEEE*, Jiakun Fang, *Senior Member, IEEE,* Jinyu Wen, *Member, IEEE,* and Yang Shi, *Fellow, IEEE*

***Abstract*—Air conditioner (AC) loads offer promising flexibility for active distribution networks to manage uncertainties, such as those in renewable energy generation, electricity prices, and load demand. However, real-time scheduling of ACs is challenging due to their massive temporal coupling constraints and time-causal uncertainties. To address this, a novel time-causal aggregation-based approximate dynamic programming (TCA-ADP) algorithm is proposed for efficient scheduling. The time-causality requirements for aggregating state variables are first analyzed to align with the real-time sequential decision-making process. Subsequently, an enhanced aggregation model is developed to ensure both high accuracy and adherence to time causality. The aggregation process is further reformulated as a linear program to optimize aggregation parameters and enable tractable computation. Accordingly, the TCA-ADP leverages aggregated state variables to approximate the value function as a new way, balancing computational efficiency and economy against the large value function space of massive ACs. By training the value function offline using historical data, the TCA-ADP efficiently achieves near-optimal real-time scheduling of massive ACs through parallel and closed-form disaggregation. Case studies demonstrate the effectiveness and scalability of the TCA-ADP, highlighting its aggregation accuracy, uncertainty handling, and the trade-off between economy and tractability.**

## NOMENCLATURE

In this paper, main symbols and notations are listed here. Others will be defined as required. Also, we use italics for single parameter or variable, bold fonts for matrices or vectors, and calligraphic fonts for sets.

*A. Abbreviations:*

| | |
|---|---|
| ADN | Active distribution network. |
| AC | Air conditioner. |
| MDP | Markov decision process. |
| ADP | Approximate dynamic programming. |
| TCA-ADP | Time-causal aggregation-based ADP. |
| MPC | Model predictive control. |
| DG | Distributed generator. |
| WT | Wind turbine. |
| PV | Photovoltaic panel. |
| EG | External grid. |

*B. Indices and Sets:*

| | |
|---|---|
| $t/N^T$ | Period index/number. |
| $i/j$ | Bus index. |
| $k/N^K$ | AC index/number at one bus. |
| $s/N^S$ | Segment index/number. |
| $\kappa/N^\kappa$ | Iteration index/number. |
| $\mathbb{U}^{AC}$ | Individual AC set. |
| $\mathbb{U}^{agg}$ | Exact aggregate AC set. |
| $\mathbb{P}^{agg}$ | Approximate aggregate AC set. |
| $\mathbb{U}^{base}$ | Base set for ACs. |
| $\mathbb{R}$ | Real number set. |

*C. Parameters:*

| | |
|---|---|
| $c^{trans}$ | Prices of power transaction with EG. |
| $c^{dg}$ | Prices of DG generation. |
| $c^{dr}$ | Prices of AC response. |
| $\gamma$ | Power factor angle of ACs. |
| $P^{ref}$ | Reference power of ACs. |
| $\Delta t$ | Period length. |
| $a^{c1}/\dots/a^{c7}$ | Thermal coefficients of ACs. |
| $T^o$ | Outdoor temperature. |
| $H^a/H^m$ | Thermal conductance of indoor air/mass. |
| $C^a/C^m$ | Thermal capacity of indoor air/mass. |
| $\mu^h/f^h$ | Coefficients for ACs in [0,1]. |
| $T^{set}/\beta^{set}$ | Indoor temperature set-point/tolerance. |
| $\underline{P^a}/\overline{P^a}$ | Minimum/Maximum $P^a$. |
| $P^{ld,a}/Q^{ld,a}$ | Forecast active/reactive load demand. |
| $P^{pv,a}/P^{wd,a}$ | Forecast available PV/WT power. |
| $\boldsymbol{G}$ | Real part of admittance matrix. |
| $\boldsymbol{B}/\boldsymbol{B}'$ | Imaginary part of admittance matrix with/ without shunt elements. |
| $g/b$ | Branch conductance/susceptance. |
| $\underline{V}/\overline{V}$ | Minimum/Maximum $V$. |
| $\overline{P^L}$ | Maximum $P^L$. |
| $\underline{P^{dg}}/\overline{P^{dg}}$ | Minimum/Maximum $P^{dg}$. |
| $\underline{Q^{dg}}/\overline{Q^{dg}}$ | Minimum/Maximum $Q^{dg}$. |
| $\underline{Q^{wd}}/\overline{Q^{wd}}$ | Minimum/Maximum $Q^{wd}$. |
| $\underline{Q^{pv}}/\overline{Q^{pv}}$ | Minimum/Maximum $Q^{pv}$. |
| $\overline{P^{eg}}/\overline{Q^{eg}}$ | Maximum $P^{eg}/Q^{eg}$. |
| $\boldsymbol{\Gamma}^{aff}$ | Affine matrix for aggregation. |
| $\boldsymbol{\gamma}^{aff}$ | Translation vector for aggregation. |

This work was supported by the National Natural Science Foundation of China (52177088 and 52207108). *(Corresponding author: Xiaomeng Ai).*

Jingguan Liu, Xiaomeng Ai, Shichang Cui, Xizhen Xue, Shengshi Wang, Jiakun Fang, and Jinyu Wen are with the State Key Laboratory of Advanced Electromagnetic Technology, Huazhong University of Science and Technology, Wuhan 430074, China (e-mail: xiaomengai@hust.edu.cn).

Xizhen Xue is also with the School of Electrical and Electronic Engineering, Nanyang Technological University, 639798, Singapore.

Yang Shi is with the Department of Mechanical Engineering, University of Victoria, Victoria, BC V8W 3P1, Canada.

| | |
|---|---|
| $r^a/r^m$ | Segment slopes for $T^{a,agg}/T^{m,agg}$. |

*D. Decision Variables:*

| | |
|---|---|
| $C^{obj}$ | Scheduling cost. |
| $P^{buy}/P^{sell}$ | Power purchased/sold from EG. |
| $P^{eg}/Q^{eg}$ | Active/Reactive power injected from EG. |
| $P^{dg}/Q^{dg}$ | Active/Reactive power output of DG. |
| $P^{dru}/P^{drd}$ | Up/down response power of ACs. |
| $P^{agg}/Q^{agg}$ | Active/Reactive power of aggregate ACs. |
| $P^a$ | Cooling power of individual AC. |
| $T^a/T^m$ | Indoor air/mass temperature. |
| $P^{wd}/Q^{wd}$ | Active/Reactive power output of WT. |
| $P^{pv}/Q^{pv}$ | Active/Reactive power output of PV. |
| $V/\theta$ | Bus voltage magnitude/phase angle. |
| $P^L$ | Active power flow. |
| $v^a/v^m$ | Segment length of $T^{a,agg}$ and $T^{m,agg}$. |

## I. Introduction

THE traditional distribution network is transitioning to an active distribution network (ADN) with the proliferation of renewables [1]. Due to the intermittency and uncertainty of renewables, ADN requires more flexible resources to follow net load changes in real-time scheduling [2]. Meanwhile, as a major energy consumer in ADNs, particularly during the summer, air conditioning (AC) loads represent a promising flexible resource on the demand side [3]. In recent years, inverter ACs, known for their higher energy efficiency, have been gaining market share over traditional on/off fixed-frequency ACs [4]. Unlike fixed-frequency units, inverter ACs can precisely match the set temperature by continuously adjusting the compressor frequency, offering greater flexibility in cooling power regulation [5]. This thermal inertia and adjustable operating range allow inverter ACs to function as virtual energy storage systems, providing significant flexibility to ADNs and enhancing the economic performance of system scheduling under uncertainty [6]. In this regard, this paper focuses on the use of inverter ACs as a key flexible resource.

The real-time scheduling of ADN integrated with ACs and uncertainty is critical, aiming to minimize operation costs while satisfying massive device constraints [7]. However, since the thermal dynamics of ACs are originally described by time-domain differential equations, their feasible region exhibits massive strong temporal coupling constraints [8]. Thus, the main difficulty lies in handling the inter-period characteristics of massive ACs under any realization of uncertainty in each scheduling period.

Extensive scheduling algorithms have been developed for ACs under uncertainty, including worst-case scenario-based robust optimization [9], typical-case scenario-based stochastic programming [10], worst-distribution-based distributionally robust optimization [11], risk-preference-based chance constraints [7], distributionally robust chance constraints considering both robustness and risk preference [12], and etc. While these algorithms are effective in handling uncertainty, their applicability to real-time scheduling may be limited. This limitation arises because they typically require a global realization of uncertainty to derive scheduling results, which violates the time causality in real-time scheduling (a.k.a. *non-anticipativity* in the literature). In fact, the real-time scheduling is a multi-period sequential decision process where decisions are made period-by-period [13]. That is, decisions made in a certain period only depend on the currently observed information rather than the information realized in future periods. Ignoring time causality can severely underestimate the effect of uncertainty, leading to poor or even infeasible scheduling [14]. Therefore, ensuring time causality is crucial to meet the needs of real-time applications.

In response to this, some recent studies have sought to address time-causal uncertainty in decision-making. In [15], multi-stage robust optimization is implemented using explicit and implicit decision algorithms to ensure time causality. However, the utilization of probability distribution knowledge from historical data is not fully exploited, leading to overly conservative scheduling results [16]. In [17], a multi-stage stochastic dual dynamic programming algorithm is developed, but its solution quality is achieved at the expense of significant computational burden, limiting its applicability. In [18], the model predictive control (MPC) algorithm is applied for real-time scheduling with continuously updated information, allowing for more practical decisions to be made into the future. However, MPC heavily relies on the accuracy of the prediction. If real-time forecast information is inaccurate or missing, the application of MPC will face significant challenges.

To overcome these limitations, the emerging approximate dynamic programming (ADP) algorithm presents a promising solution. It employs Bellman's equation to decompose the multi-period problem into several single-period subproblems, which are then sequentially solved over time [19]. In order to ensure solution quality under uncertainty, ADP leverages offline learning of empirical knowledge from historical data to train value functions. Subsequently, it acquires near-optimal real-time scheduling results online with the aid of pre-trained value functions. As a result, ADP has demonstrated successful applications in battery energy storage operation [20], unit dispatch [21], multi-energy systems [16], and other domains.

However, the existing ADP algorithms mentioned above are primarily tailored for small-scale storage-like assets with a limited number of state variables. Consequently, the dimensionality of the value function space is often constrained, allowing for convenient training of well-fitted value functions. Consequently, they may not be suitable for real-time scheduling of massive ACs. Given that ACs are characterized by small individual capacities but a large overall quantity, the integration of massive ACs to meet the entry thresholds of ADN introduces significant computational complexity [22]. This means that the presence of massive ACs results in an exceedingly extensive value function space, leading to an unacceptably long training time for value function fitting. Thus, existing ADP algorithms get stuck in this context.

To bridge this research gap, we introduce a time-causal aggregation-based approximate dynamic programming (TCA-ADP) algorithm for the efficient real-time scheduling of massive ACs under time-casual uncertainty. Compared to the state-of-art, our main contributions are twofold:

1) *New Insight*: To our knowledge, this is the first analysis of time causality requirements for aggregating state variables, aligning them with practical real-time decision-making processes. Given this, an enhanced state variable aggregation model is designed, ensuring high aggregation accuracy and time causality. To optimize the aggregation parameters, we reformulate the aggregation process as a linear program for tractable computation.

2) *New Algorithm*: A TCA-ADP algorithm is adapted for efficient scheduling. It leverages time-causal state variable aggregation to fit the value function as a new way, balancing computational efficiency and economy against the large value function space of massive ACs in real-time scheduling. By training the value function offline around aggregated state variables using historical data, the proposed TCA-ADP efficiently achieves near-optimal real-time scheduling for massive ACs via parallel and closed-form disaggregation.

Note that for other flexible resources whose feasible regions can be expressed or approximated by convex polytopes in half-space representation, the proposed state variable aggregation is still applicable. With this in mind, while this paper focuses on the real-time scheduling of massive ACs, the proposed TCA-ADP algorithm can also be adapted with minor adjustments to schedule other types of large-scale flexible resources with similar inter-period characteristics, such as energy storages [23], HVAC systems [24], and electric vehicles [25].

The remainder of this paper is organized as follows. Section II presents the problem formulation in real-time scheduling. Section III discusses the TCA-ADP scheduling algorithm. Case studies are conducted in Section IV and conclusions are drawn in Section V.

## II. Problem Formulation

The ADN studied herein includes distributed generators (DGs), the external grid (EG), wind turbines (WTs), photovoltaic panels (PVs), massive flexible ACs, and inelastic loads. The real-time scheduling model is formulated as a multi-period ($N^T$ periods) sequential decision-making problem (1.a)-(1.h) where the decisions in period $t$ are made solely based on the available information in period $t$, but not on the future observations. The objective function (1.a) minimizes the total expected scheduling cost $\mathbb{E}\{\sum_t C_t^{obj}\}$ throughout the day, including power transaction cost with EG, DG generation cost, and AC response cost. Here, we assume that $c_i^{dr}$ is a pre-negotiated fixed value for direct load control, which is a common practice in user-side management [3], [26], [27]. The model is subject to aggregate AC power response constraint (1.b), individual AC operation constraint in a finite difference form (1.c)-(1.d) [8], decoupled linearized power flow constraints (1.e) [28], DG output power constraint (1.f), the active/reactive power constraint of WTs/PVs (1.g), and the exchanging power constraint with EG (1.h).

$$\min \ \mathbb{E}\left\{\sum_t C_t^{obj}\right\}, \quad C_t^{obj} = \sum_i \left( c_{i,t}^{trans}\left(P_{i,t}^{buy} - P_{i,t}^{sell}\right) + c_i^{dg} P_{i,t}^{dg} + c_i^{dr}\left(P_{i,t}^{dru} + P_{i,t}^{drd}\right)\right) \tag{1.a}$$

$$\text{s.t.} \ \begin{cases} P_{i,t}^{agg} = \sum_k P_{i,k,t}^{a}, Q_{i,t}^{agg} = P_{i,t}^{agg} \tan\gamma_i \\ P_{i,t}^{agg} = \sum_k P_{i,k,t}^{ref} + P_{i,t}^{dru} - P_{i,t}^{drd}, P_{i,t}^{dru} \ge 0, P_{i,t}^{drd} \ge 0 \end{cases} \tag{1.b}$$

$$\begin{cases} \left(T_{i,k,t+1}^{a} - T_{i,k,t}^{a}\right)/\Delta t = a_{i,k}^{c1} T_{i,k,t}^{a} + a_{i,k}^{c2} T_{i,k,t}^{m} + a_{i,k}^{c3} T_{i,k,t}^{o} + a_{i,k}^{c4} P_{i,k,t}^{a} \\ \left(T_{i,k,t+1}^{m} - T_{i,k,t}^{m}\right)/\Delta t = a_{i,k}^{c5} T_{i,k,t}^{a} + a_{i,k}^{c6} T_{i,k,t}^{m} + a_{i,k}^{c7} P_{i,k,t}^{a} \\ T_{i,k}^{set} - \beta_{i,k}^{set} \le T_{i,k,t}^{a} \le T_{i,k}^{set} + \beta_{i,k}^{set}, \underline{P_{i,k}^{a}} \le P_{i,k,t}^{a} \le \overline{P_{i,k}^{a}} \end{cases} \tag{1.c}$$

$$\begin{cases} a_{i,k}^{c1} = -\left(H_{i,k}^{a} + H_{i,k}^{m}\right)/C_{i,k}^{a}, a_{i,k}^{c2} = -H_{i,k}^{m}/C_{i,k}^{a} \\ a_{i,k}^{c3} = -H_{i,k}^{a}/C_{i,k}^{a}, a_{i,k}^{c4} = -(1-f_{i,k}^{h})\mu_{i,k}^{h}/C_{i,k}^{a} \\ a_{i,k}^{c5} = H_{i,k}^{m}/C_{i,k}^{m}, a_{i,k}^{c6} = -H_{i,k}^{m}/C_{i,k}^{m} \\ a_{i,k}^{c7} = (1-f_{i,k}^{h})\mu_{i,k}^{h}/C_{i,k}^{m} \end{cases} \tag{1.d}$$

$$\begin{cases} P_{i,t}^{eg} + P_{i,t}^{dg} + P_{i,t}^{wd} + P_{i,t}^{pv} - P_{i,t}^{ld,a} - P_{i,t}^{agg} = \sum_j G_{ij} V_{j,t} + \sum_j B'_{ij}\theta_{j,t} \\ Q_{i,t}^{eg} + Q_{i,t}^{dg} + Q_{i,t}^{wd} + Q_{i,t}^{pv} - Q_{i,t}^{ld,a} - Q_{i,t}^{agg} = -\sum_j B_{ij} V_{j,t} - \sum_j G_{ij}\theta_{j,t} \\ \underline{V_i} \le V_{i,t} \le \overline{V_i}, -\pi \le \theta_{i,t} \le \pi, -\overline{P_{ij}^{L}} \le P_{ij,t}^{L} \le \overline{P_{ij}^{L}} \\ P_{ij,t}^{L} = g_{ij}(V_{i,t} - V_{j,t}) - b_{ij}(\theta_{i,t} - \theta_{j,t}) \end{cases} \tag{1.e}$$

$$\underline{P_i^{dg}} \le P_{i,t}^{dg} \le \overline{P_i^{dg}}, \underline{Q_i^{dg}} \le Q_{i,t}^{dg} \le \overline{Q_i^{dg}} \tag{1.f}$$

$$\begin{cases} 0 \le P_{i,t}^{wd} \le P_{i,t}^{wd,a}, 0 \le P_{i,t}^{pv} \le P_{i,t}^{pv,a} \\ \underline{Q_i^{wd}} \le Q_{i,t}^{wd} \le \overline{Q_i^{wd}}, \underline{Q_i^{pv}} \le Q_{i,t}^{pv} \le \overline{Q_i^{pv}} \end{cases} \tag{1.g}$$

$$\begin{cases} P_{i,t}^{eg} = P_{i,t}^{buy} - P_{i,t}^{sell}, P_{i,t}^{buy} \ge 0, P_{i,t}^{sell} \ge 0 \\ -\overline{P_i^{eg}} \le P_{i,t}^{eg} \le \overline{P_i^{eg}}, -\overline{Q_i^{eg}} \le Q_{i,t}^{eg} \le \overline{Q_i^{eg}} \end{cases} \tag{1.h}$$

Note that we define the power required for an AC to maintain the indoor air temperature at the set point as the reference power $P_{i,k,t}^{ref}$ [4], which can be derived from (1.c) by setting the differential terms to zero and assuming the indoor temperature is at the set point as follows:

$$P_{i,k,t}^{ref} = \left(\left(a_{i,k}^{c6} a_{i,k}^{c1} - a_{i,k}^{c5} a_{i,k}^{c2}\right) T_{i,k}^{set} + a_{i,k}^{c6} a_{i,k}^{c3} T_{i,k,t}^{o}\right) / \left(a_{i,k}^{c6} a_{i,k}^{c4} - a_{i,k}^{c7} a_{i,k}^{c2}\right) \tag{2}$$

## III. TCA-ADP Scheduling Algorithm

This section introduces the proposed TCA-ADP scheduling algorithm, including Markov decision process (MDP) reformulation, time-causal state variable aggregation, parameter determination, value function fitting around aggregated state variables, and practical application framework.

### *A. MDP Reformulation Under Bellman Optimality Equation*

The real-time scheduling model (1.a)-(1.h) can be recast as a MDP problem by decomposing the original multi-period problem into multiple sequential single-period problems and solving them recursively. A MDP problem typically comprises four parts: state variables $\boldsymbol{S}_t$, decision variables $\boldsymbol{X}_t$, exogenous information $\widehat{\boldsymbol{W}}_t$, and state transition [16].

The state variables indicate the system's current state in period $t$, providing sufficient information about the system to determine its future behavior in the absence of any external force affecting the system. Thus, the indoor temperature variables appearing in the inter-period constraint (1.c) are regarded as state variables herein, as shown in (3).

$$\boldsymbol{S}_t = \left\{ T_{i,k,t}^{a}, T_{i,k,t}^{m} \right\} \tag{3}$$

The decision variables reflect the system actions after observing the system states in period $t$, as shown in (4) .

$$\boldsymbol{X}_t = \begin{Bmatrix} P_{i,t}^{buy}, P_{i,t}^{sell}, P_{i,t}^{eg}, Q_{i,t}^{eg}, P_{i,t}^{dg}, Q_{i,t}^{dg}, P_{i,t}^{dru}, P_{i,t}^{drd}, P_{i,k,t}^{a}, \\ P_{i,t}^{agg}, Q_{i,t}^{agg}, P_{i,t}^{wd}, Q_{i,t}^{wd}, P_{i,t}^{pv}, Q_{i,t}^{pv}, V_{j,t}, \theta_{j,t}, P_{ij,t}^{L} \end{Bmatrix} \tag{4}$$

The exogenous information represents the realization of system uncertainty sources in period $t$, including renewable outputs, active /reactive load demands, and power transaction prices as shown in (5).

$$\hat{\boldsymbol{W}}_t = \left\{ \hat{P}_{i,t}^{wd,a}, \hat{P}_{i,t}^{pv,a}, \hat{P}_{i,t}^{ld,a}, \hat{Q}_{i,t}^{ld,a}, \ \hat{c}_{i,t}^{trans} \right\} \tag{5}$$

The state transition demonstrates the state conversion process in two consecutive periods, which can be represented in a compact matrix form [13], as shown in (6). Herein, $\boldsymbol{C}^{[\cdot]}$ are compact coefficient matrices extracted from (1.a)-(1.h) where any equality can be replaced by two opposite inequalities.

$$\boldsymbol{C}^{c1}\boldsymbol{S}_{t+1} + \boldsymbol{C}^{c2}\boldsymbol{X}_t + \boldsymbol{C}^{c3}\boldsymbol{S}_t \le \boldsymbol{C}^{c4} + \boldsymbol{C}^{c5}\hat{\boldsymbol{W}}_t \tag{6}$$

Given the above MDP elements, the globally optimal solution to the multi-period problem (1.a)-(1.h) can be procured by recursively solving the single-period Bellman optimality equation (7) [20]. Here, $V_t(\boldsymbol{S}_t)$ represents the value function at state $\boldsymbol{S}_t$, indicating the optimal cost starting from that state. This value encompasses both the function value of the current state plus the cost-to-go function value. The cost-to-go function refers to the operational cost of the ADN from the subsequent scheduling period $t+1$ to the last period $N^T$.

$$V_t(\boldsymbol{S}_t) = \min \left\{ \underbrace{C_t^{obj}(\boldsymbol{S}_t, \boldsymbol{X}_t)}_{\text{current cost}} + \underbrace{\mathbb{E}[V_{t+1}(\boldsymbol{S}_{t+1}) \mid \boldsymbol{S}_t]}_{\text{cost-to-go function}} \right\} \tag{7}$$

The classical dynamic programming method first solves Bellman's equations (7) in reverse time to recursively compute the value functions for all possible states. It then derives the optimal solutions by solving the Bellman's equations forward in time using the computed value functions. However, in stochastic environments, evaluating all possible uncertainty scenarios in real-time to compute the expectation in (7) is computationally impractical. This is known as the "curse of dimensionality" in the traditional dynamic programming method, which renders it ineffective for solving problem (7).

To address this, ADP methodologies, as outlined in [20], [29], [30], replace the expectation in (7) with the post-decision value function $V_t^x(\boldsymbol{S}_t^x)$, approximated around the post-decision state $\boldsymbol{S}_t^x$ to quantify the influence of current decisions on future costs. The post-decision state $\boldsymbol{S}_t^x$ represents the system state soon after the decisions in period $t$ are made, but before any uncertainty information in period $t+1$ is observed. By using post-decision state variables, the optimization over high-dimensional random factors is avoided, reducing computational complexity while providing accurate approximations of the original expected values [31]. This substitution allows the Bellman equation to be reformulated using the post-decision value function, as shown in (8).

$$V_t(\boldsymbol{S}_t) \simeq \min \left\{ C_t^{obj}(\boldsymbol{S}_t, \boldsymbol{X}_t) + V_t^x(\boldsymbol{S}_t^x) \right\} \tag{8}$$

In summary, the sequential decision process of the MDP problem is illustrated in Fig. 1. In period $t$, the current state $\boldsymbol{S}_t$ is already known, and $\widehat{\boldsymbol{W}}_t$ is observed. The decision-maker obtains the optimized value of decision variables $\boldsymbol{X}_t^*$ by solving (9), derives the post-decision state $\boldsymbol{S}_t^x$, and moves to the next state $\boldsymbol{S}_{t+1}$ via state transition (6). Then, this process is repeated period-by-period for subsequent periods.

$$\boldsymbol{X}_t^* = \arg\min \left\{ C_t^{obj}(\boldsymbol{S}_t, \boldsymbol{X}_t) + V_t^x(\boldsymbol{S}_t^x) \right\} \tag{9}$$

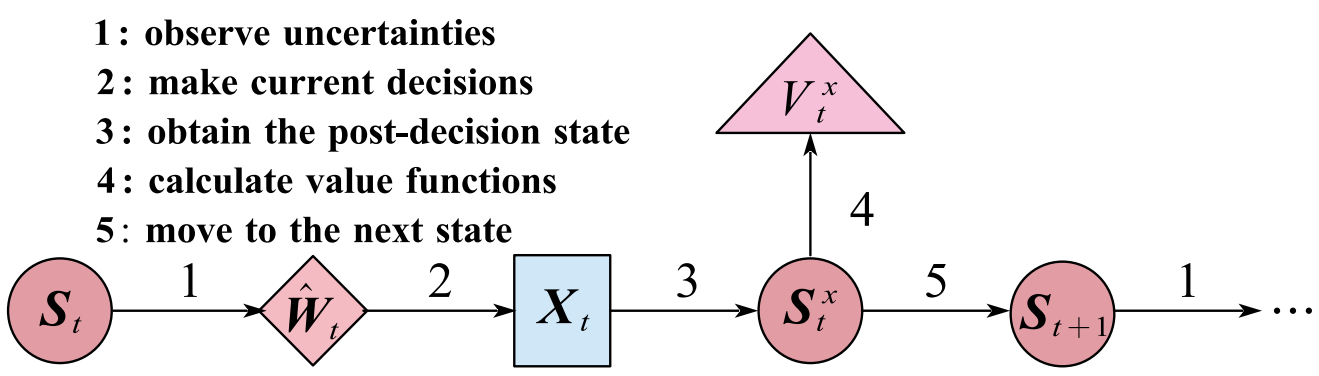

Fig. 1. Sequential decision process of the MDP problem.

*B. Time-Causal State Variable Aggregation Model*

To ensure optimality, it is necessary to properly fit the post-decision value functions to map $\boldsymbol{S}_t^x$ to $V_t^x$. However, the existence of massive state variables from ACs results in an exceedingly large value function space, posing challenges for value function fitting. To resolve this, the time-causal state variable aggregation is proposed.

We first recast the individual AC operation constraint (1.c) as a flexibility set in the H-representation form, as shown in (10). The bold variable vector represents a collection of corresponding italic variables arranged in time sequence, for example, $\boldsymbol{P}_{i,k}^a = \left[ P_{i,k,1}^a, P_{i,k,2}^a, \dots, P_{i,k,N^T}^a \right]^{\mathrm{T}}$. $\boldsymbol{H}$ / $\boldsymbol{h}$ is coefficient matrix/vector parameter of individual AC sets, which can be derived from (1.c).

$$\mathbb{U}_{i,k}^{AC} = \left\{ \boldsymbol{H}_{i,k} \left[ \left(\boldsymbol{P}_{i,k}^a\right)^{\mathrm{T}}, \left(\boldsymbol{T}_{i,k}^a\right)^{\mathrm{T}}, \left(\boldsymbol{T}_{i,k}^m\right)^{\mathrm{T}} \right]^{\mathrm{T}} \le \boldsymbol{h}_{i,k} \right\} \tag{10}$$

For $N_i^K$ ACs within node $i$, the aggregate flexibility set can be expressed as a Minkowski sum of the individual AC sets [32], as shown in (11) where $\uplus$ is the Minkowski sum calculation.

$$\mathbb{U}_i^{agg} = \biguplus_k \mathbb{U}_{i,k}^{AC} = \left\{ \boldsymbol{P}_i^{agg} = \sum_k \boldsymbol{P}_{i,k}^a, \boldsymbol{P}_{i,k}^a \in \mathbb{U}_{i,k}^{AC} \right\} \tag{11}$$

While the Minkowski sum is a classical problem in computational geometry, it is generally NP-hard [33]. Given this, our primary goal is to develop a state variable aggregation model that approximates the aggregate flexibility of ACs and, most importantly, is suitable for real-time scheduling. Specifically, we aim to identify an inner approximate set $\mathbb{P}_i^{agg}$:

$$\mathbb{P}_i^{agg} \subseteq \mathbb{U}_i^{agg} \tag{12}$$

*1) Time-Causality Requirement:* We first dissect the time-causality requirements for state variable aggregation. Considering the sequential decision-making process shown in Fig. 1 and the temporal coupling characteristics in individual AC sets, the left-hand-side coefficient matrix $\boldsymbol{H}_{i,k}$ of the individual AC set $\mathbb{U}_{i,k}^{AC}$ can be constructed using matrix blocks in the following form:

$$\boldsymbol{H}_{i,k}=\begin{bmatrix}\boldsymbol{A}_{11} & \cdots & \boldsymbol{A}_{1q}\\ \vdots & \vdots & \vdots\\ \boldsymbol{A}_{p1} & \cdots & \boldsymbol{A}_{pq}\end{bmatrix} \tag{13.a}$$

$$\boldsymbol{A}_{pq}=\begin{bmatrix}a_{11} & 0 & 0 & \cdots & 0 & 0\\ a_{21} & a_{22} & 0 & \cdots & 0 & 0\\ 0 & a_{32} & a_{33} & \cdots & 0 & 0\\ \vdots & \vdots & \vdots & \cdots & \vdots & \vdots\\ 0 & 0 & 0 & \cdots & a_{(N^T-1)(N^T-1)} & 0\\ 0 & 0 & 0 & \cdots & a_{N^T(N^T-1)} & a_{N^TN^T}\end{bmatrix} \tag{13.b}$$

where $p$ and $q$ are the indices of row and column respectively. Only the element $a$ in $\boldsymbol{A}_{pq}$ can be non-zero.

Intuitively, the structure of $\boldsymbol{A}_{pq}$ means that the constraints of individual AC set in period $t$ only comprise the variables in current period $t$ and next period $t+1$. That is, the individual AC sets can be recast as a MDP for real-time sequential decision process. In this regard, we say that a set is *time-causal* if its left-hand side coefficient matrix can be represented in the form of (13.a)-(13.b). Also, we say a matrix is a *time-causal matrix* if the matrix has the same structure as $\boldsymbol{A}_{pq}$.

According to the structure of (10), the individual AC sets are naturally time-causal. Hence, the exact aggregate set $\mathbb{U}_i^{agg}$ is also time-causal. In light of this, we need to ensure the approximate set $\mathbb{P}_i^{agg}$ is also time-causal for real-time implementation. With this task in mind, we then propose the state variable aggregation model to derive $\mathbb{P}_i^{agg}$.

*2) Aggregation Model:* A pioneering method for the inner approximation of the Minkowski sum was introduced in [32], where the homothet of a given template polyhedron is utilized for efficient Minkowski sum calculation, and the inner approximation ensures disaggregation feasibility. However, this method has been observed to sacrifice much flexibility due to its poor geometric adaptability [25]. To resolve this, we adopt the idea in [32], but enhance the geometric adaptability by improving the geometric transformation technique from uniform scaling in [32] to a general affine transformation.

Three key steps are proposed to efficiently derive $\mathbb{P}_i^{agg}$:

Step 1: A base set $\mathbb{U}_i^{base}$ is constructed as shown in (14.a). The coefficient matrix $\boldsymbol{H}_i^{base}$ and vector $\boldsymbol{h}_i^{base}$ are predefined known parameters, which can be specified via averaging from all ACs within node $i$. $\boldsymbol{P}_i^{a,base}/\boldsymbol{T}_i^{a,base}/\boldsymbol{T}_i^{m,base}$ represents the variables in $\mathbb{U}_i^{base}$ sharing the same dimension as $\boldsymbol{P}_{i,k}^{a}/\boldsymbol{T}_{i,k}^{a}/\boldsymbol{T}_{i,k}^{m}$.

Step 2: For each AC, a diagonal affine matrix parameter $\boldsymbol{\Gamma}_{i,k}^{aff}\in\mathbb{R}^{N^T\times N^T}$ and a translation vector parameter $\boldsymbol{\gamma}_{i,k}^{aff}\in\mathbb{R}^{N^T}$ are carefully determined such that the affine-transformed base set closely approximates the original AC set $\mathbb{U}_{i,k}^{AC}$ while respecting the inner containment constraints (14.b).

Step 3: The approximate aggregate set $\mathbb{P}_i^{agg}$ can be efficiently calculated in a closed form using (14.c).

$$\mathbb{U}_i^{base}=\left\{\boldsymbol{H}_i^{base}\left[\left(\boldsymbol{P}_i^{a,base}\right)^{\mathrm{T}},\left(\boldsymbol{T}_i^{a,base}\right)^{\mathrm{T}},\left(\boldsymbol{T}_i^{m,base}\right)^{\mathrm{T}}\right]^{\mathrm{T}}\le\boldsymbol{h}_i^{base}\right\} \tag{14.a}$$

$$\boldsymbol{\Gamma}_{i,k}^{aff}\mathbb{U}_i^{base}+\boldsymbol{\gamma}_{i,k}^{aff}\subseteq\mathbb{U}_{i,k}^{AC},\ \boldsymbol{\Gamma}_{i,k}^{aff}\ \text{is a diagonal matrix} \tag{14.b}$$

$$\mathbb{P}_i^{agg}=\sum_k\left(\boldsymbol{\Gamma}_{i,k}^{aff}\right)\mathbb{U}_i^{base}+\sum_k\left(\boldsymbol{\gamma}_{i,k}^{aff}\right)\subseteq\mathbb{U}_i^{agg} \tag{14.c}$$

Let $\boldsymbol{P}_i^{agg}\in\mathbb{P}_i^{agg}$ denote an arbitrary point in $\mathbb{P}_i^{agg}$. Since $\mathbb{P}_i^{agg}=\sum_k(\boldsymbol{\Gamma}_{i,k}^{aff})\,\mathbb{U}_i^{base}+\sum_k(\boldsymbol{\gamma}_{i,k}^{aff})$, there exists a point $\boldsymbol{P}_i^{a,base}\in\mathbb{U}_i^{base}$ such that the given point $\boldsymbol{P}_i^{a,base}$ can be expressed as (15.a).

$$\boldsymbol{P}_i^{agg}=\left(\sum_k(\boldsymbol{\Gamma}_{i,k}^{aff})\right)\boldsymbol{P}_i^{a,base}+\left(\sum_k(\boldsymbol{\gamma}_{i,k}^{aff})\right) \tag{15.a}$$

Combining (14.a) and (15.a), we can explicitly express $\mathbb{P}_i^{agg}$ as (15.b), where $\boldsymbol{P}_i^{agg}/\boldsymbol{T}_i^{a,agg}/\boldsymbol{T}_i^{m,agg}$ represents the aggregated variable and shares the same dimension as $\boldsymbol{P}_{i,k}^{a}/\boldsymbol{T}_{i,k}^{a}/\boldsymbol{T}_{i,k}^{m}$. Note that details regarding the invertibility of matrix $\sum_k(\boldsymbol{\Gamma}_{i,k}^{aff})$ are provided in Appendix D.

$$\mathbb{P}_i^{agg}=\left\{\boldsymbol{H}_i^{base}[\left(\sum_k\boldsymbol{\Gamma}_{i,k}^{aff}\right)^{-1}\left(\boldsymbol{P}_i^{agg}-\sum_k\boldsymbol{\gamma}_{i,k}^{aff}\right),\boldsymbol{T}_i^{a,agg},\boldsymbol{T}_i^{m,agg}]^{\mathrm{T}}\le\boldsymbol{h}_i^{base}\right\} \tag{15.b}$$

*3) An Illustrative Toy Example:* To demonstrate the features of the proposed state variable aggregation, we approximate a three-period time-causal set for visualization, which involves three state variables $S_1\sim S_3$. The coefficient matrices/vectors of the time-causal exact set and base set are provided in Appendix A. We compare the following three approximation methods and show the results in Fig. 2 and TABLE I.

**M1**: Using the traditional method, i.e., the uniform scaling factor, for approximation [32].

**M2(Proposed)**: Using a diagonal affine matrix for approximation.

**M3**: Using a dense affine matrix for approximation, no longer constraining the affine matrix to be a diagonal matrix.

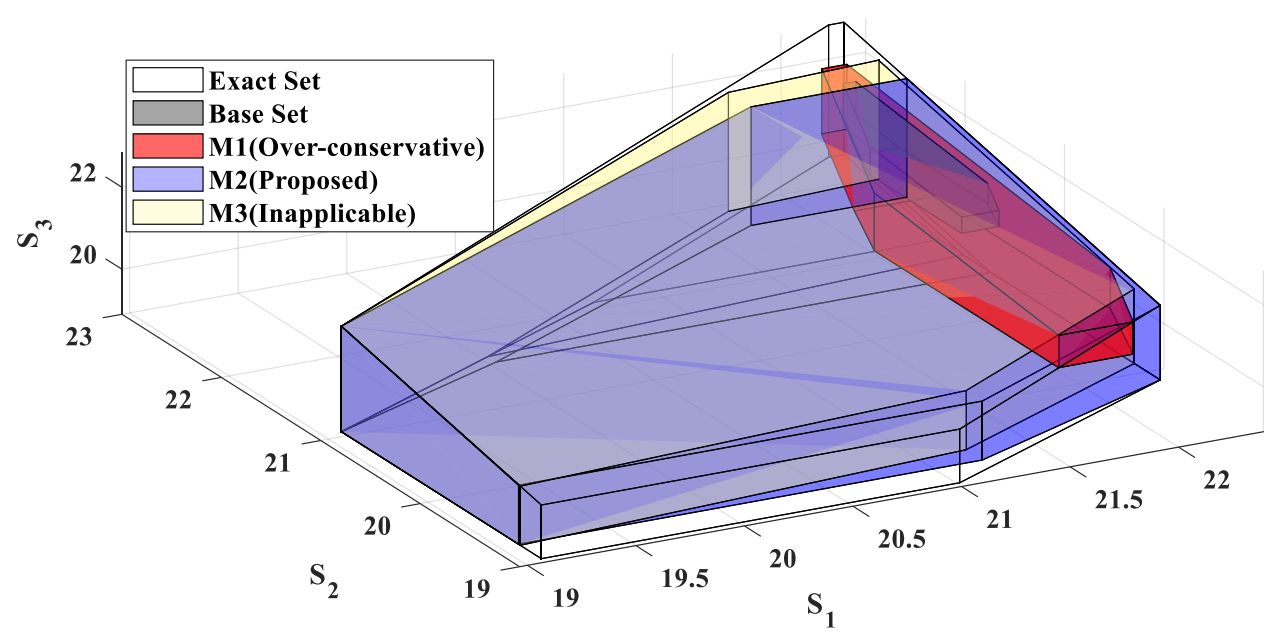

Fig. 2. Approximation results in the illustrative example.

TABLE I
APPROXIMATION RESULTS OF THE ILLUSTRATIVE EXAMPLE

| Item | Exact Set | M1 | M2 | M3 |
|---|---|---|---|---|
| Transformation Matrix $\boldsymbol{\Gamma}^{aff}$ | / | $\begin{bmatrix}2&0&0\\0&2&0\\0&0&2\end{bmatrix}$ | $\begin{bmatrix}12.44&0&0\\0&1.92&0\\0&0&3.65\end{bmatrix}$ | $\begin{bmatrix}12.44&0&0\\1.05&1.94&0\\0&0&3.66\end{bmatrix}$ |
| Time Causality of Set | **Yes** | **Yes** | **Yes** | **No** |
| Set Volume | 21.50 | 1.70 | 18.53 | 18.76 |

From the above results, we observe that the traditional **M1** yields an over-conservative approximation with unacceptable flexibility loss (smallest volume). On the other hand, **M3** demonstrates the most accurate approximation (large volume close to exact set), but it violates time causality. Specifically, one constraint of the approximate set in **M3** is $0.4S_1 - 4.76S_2 + 7.55S_3 \leqslant 70.61$. As a result, the state transition from $S_1$ to $S_2$ in period 1 depends on the unrevealed state variable $S_3$ in the future period, leading to *a failure of state transition*. Thus, **M3** is not suitable for real-time scheduling. In contrast, the proposed **M2** provides a much more accurate approximation than **M1** (much larger volume) due to its enhanced geometric adaptability. Also, the proposed **M2** maintains its approximate set as a time-causal set by constraining $\boldsymbol{\Gamma}_{i,k}^{aff}$ to be a diagonal matrix. This feature can be explained from two perspectives:

*(a) Geometric perspective*: The proposed **M2** only stretches the base set along different dimensions, without altering the correlation among the different dimensions of state variables.

*(b) Algebraic perspective*: Multiplying a time-causal matrix by a diagonal matrix still results in a time-causal matrix. Thus, the time causality is preserved in the proposed **M2**.

In summary, the proposed state variable aggregation model balances aggregation accuracy and time causality requirements, making it suitable for practical sequential scheduling.

### *C. Parameter Determination for Aggregation Model*

*1) Optimization Problem for Parameter Determination:* The remaining task is to determine the optimal value of $\boldsymbol{\Gamma}_{i,k}^{aff}$ and $\boldsymbol{\gamma}_{i,k}^{aff}$ such that $\boldsymbol{\Gamma}_{i,k}^{aff}\mathbb{U}_i^{base} + \boldsymbol{\gamma}_{i,k}^{aff}$ forms a volume-maximization subset of $\mathbb{U}_{i,k}^{AC}$. To achieve this objective, we formulate the following optimization problem for each AC.

$$\max_{\boldsymbol{\Gamma}_{i,k}^{aff},\boldsymbol{\gamma}_{i,k}^{aff}} \det\left(\boldsymbol{\Gamma}_{i,k}^{aff}\right) \tag{16.a}$$

$$\text{s.t. } \boldsymbol{\Gamma}_{i,k}^{aff}\mathbb{U}_i^{base} + \boldsymbol{\gamma}_{i,k}^{aff} \subseteq \mathbb{U}_{i,k}^{AC} \tag{16.b}$$

$$\boldsymbol{\Gamma}_{i,k}^{aff} \text{ is a diagonal matrix} \tag{16.c}$$

Here, the objective function (16.a) maximizes the determinant of $\boldsymbol{\Gamma}_{i,k}^{aff}$, ensuring the largest inner approximation. The constraint (16.b) guarantees the containment relationship for inner approximation, while constraint (16.c) limits the form of $\boldsymbol{\Gamma}_{i,k}^{aff}$.

However, two nonlinear terms should be addressed to efficiently solve the problem (16.a)-(16.c): the objective function (16.a) and the containment constraint (16.b). Thus, we will discuss the customized linear reformulation for tractable calculation.

*2) Objective Function Reformulation:* Let $\lambda_{i,k,m}^{aff}$ represents the $m-th$ diagonal elements of $\boldsymbol{\Gamma}_{i,k}^{aff}$, then we can use $\prod_m \lambda_{i,k,m}^{aff}$ to represent $\det(\boldsymbol{\Gamma}_{i,k}^{aff})$. Because the logarithmic function is strictly increasing, we can equivalently recast (16.a) as (17).

$$\max_{\boldsymbol{\Gamma}_{i,k}^{aff},\boldsymbol{\gamma}_{i,k}^{aff}} \sum_m \ln \lambda_{i,k,m}^{aff} \tag{17}$$

Further, we build a piecewise approximation for (17) to eliminate non-linearity, as shown in Fig. 3. There are $N^D$ sample points $\lambda_{i,k,m,d}^{aff,sam}, d = 1{:}N^D$, then the tangents at sample points can be expressed by (18).

$$y_{i,k,m}^{aff} - \ln \lambda_{i,k,m,d}^{aff,sam} = \left(\lambda_{i,k,m}^{aff} - \lambda_{i,k,m,d}^{aff,sam}\right) / \lambda_{i,k,m,d}^{aff,sam} \tag{18}$$

Thus, we can approximate the original logarithm function using piecewise linear function composed of $N^D$ tangents. Further, we can approximate the objective function (17) by introducing auxiliary variables $e_{i,k,m}^{aff}$ as shown in (19.a)-(19.b).

$$\max_{\boldsymbol{\Gamma}_{i,k}^{aff},\boldsymbol{\gamma}_{i,k}^{aff},e_{i,k,m}^{aff}} \sum_m e_{i,k,m}^{aff} \tag{19.a}$$

$$e_{i,k,m}^{aff} \le \left(\lambda_{i,k,m}^{aff} - \lambda_{i,k,m,d}^{aff,sam}\right) / \lambda_{i,k,m,d}^{aff,sam} + \ln \lambda_{i,k,m,d}^{aff,sam} \tag{19.b}$$

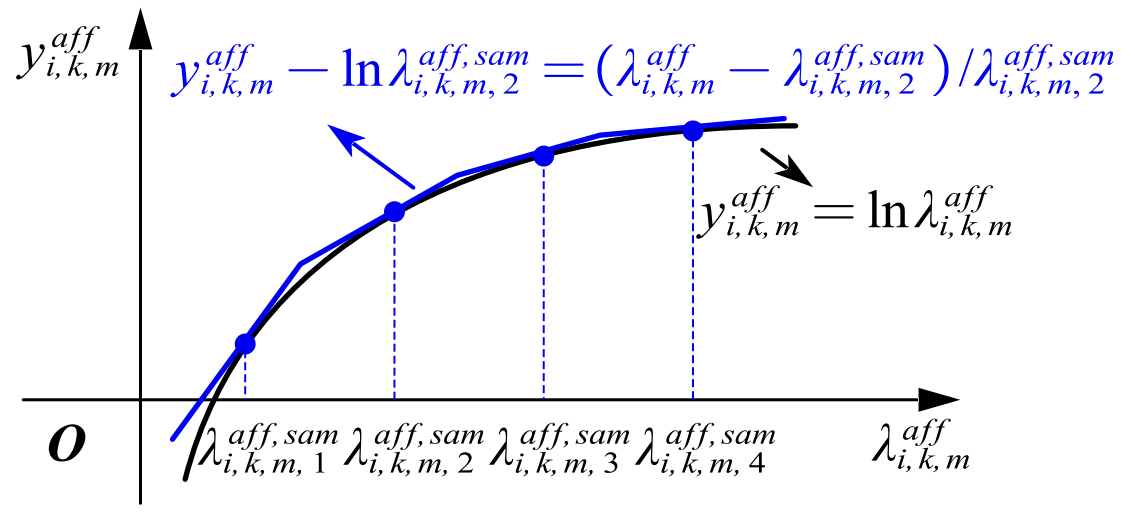

Fig. 3. Piecewise approximation for linearization.

*3) Containment Constraint Reformulation:* The existence of equality constraints in (1.c) renders the AC set not a full-dimensional region, leading to a dimension missing issue in aggregation [25]. To resolve this, we recast the AC set (10) as a full-dimensional region in power space, as shown in (20), by removing redundant variables based on equality constraints. Here, $\boldsymbol{H}^{proj}/\boldsymbol{h}^{proj}$ is known matrix/vector parameter for $\mathbb{U}_{i,k}^{AC}$. The detailed removing process can be found in Appendix B.

$$\mathbb{U}_{i,k}^{AC} = \left\{\boldsymbol{H}_{i,k}^{proj}\boldsymbol{P}_{i,k}^{a} \le \boldsymbol{h}_{i,k}^{proj}\right\} \tag{20}$$

Similarly, the base set (14.a) can also be recast as a full-dimensional region, as shown in (21). Here, $\boldsymbol{H}^{base,proj}$ /$\boldsymbol{h}^{base,proj}$ is known matrix/vector parameter for $\mathbb{U}_i^{base}$

$$\mathbb{U}_i^{base} = \left\{\boldsymbol{H}_i^{base,proj}\boldsymbol{P}_{i,k}^{a} \le \boldsymbol{h}_i^{base,proj}\right\} \tag{21}$$

After the removing process, the dimension missing issue is avoided, and the containment constraint (16.b) can be recast as shown in (22).

$$\boldsymbol{\Gamma}_{i,k}^{aff}\boldsymbol{P}_{i,k}^{a} + \boldsymbol{\gamma}_{i,k}^{aff} \subseteq \mathbb{U}_{i,k}^{AC}, \forall \boldsymbol{P}_{i,k}^{a} \in \mathbb{U}_i^{base} \tag{22}$$

Then, we reformulate (22) as tractable linear constraints (23.a)-(23.c) and prove the reformulation correctness in Appendix C.

$$\boldsymbol{\Lambda}_{i,k} \ge 0 \tag{23.a}$$

$$\boldsymbol{\Lambda}_{i,k}\boldsymbol{H}_i^{base,proj} = \boldsymbol{H}_{i,k}^{proj}\boldsymbol{\Gamma}_{i,k}^{aff} \tag{23.b}$$

$$\boldsymbol{\Lambda}_{i,k}\boldsymbol{h}_i^{base,proj} + \boldsymbol{H}_{i,k}^{proj}\boldsymbol{\gamma}_{i,k}^{aff} \le \boldsymbol{h}_{i,k}^{proj} \tag{23.c}$$

where $\boldsymbol{\Lambda}_{i,k}$ is the auxiliary variable matrix.

*4) Reformulated Problem*: Given the known parameters of ACs, including $\boldsymbol{H}_{i,k}^{proj}$ and $\boldsymbol{h}_{i,k}^{proj}$, as well as the known parameters of the base set, $\boldsymbol{H}_i^{base,proj}$ and $\boldsymbol{h}_i^{base,proj}$, we can treat $\boldsymbol{\Gamma}_{i,k}^{aff}$, $\boldsymbol{\gamma}_{i,k}^{aff}$, $\boldsymbol{\Lambda}_{i,k}$, and $e_{i,k,m}^{aff}$ as decision variables to reformulate the original optimization problem (16.a)-(16.c) as a linear program (24.a)-(24.b), which can be solved by off-the-shelf solvers. After solving this program, we can derive the optimal value of $\boldsymbol{\Gamma}_{i,k}^{aff}$ and $\boldsymbol{\gamma}_{i,k}^{aff}$ to form $\mathbb{P}_i^{agg}$.

$$\max_{\boldsymbol{\Gamma}_{i,k}^{aff},\boldsymbol{\gamma}_{i,k}^{aff},e_{i,k,m}^{aff},\boldsymbol{\Lambda}_{i,k}} \sum_m e_{i,k,m}^{aff} \tag{24.a}$$

$$\text{s.t. (16.c),(19.b),(23.a)-(23.c)} \tag{24.b}$$

*D. Value Function Fitting Around Aggregated State Variables*

By leveraging the proposed state variable aggregation, the original constrains (1.c)-(1.d) of massive ACs in the scheduling model can be replaced by $\mathbb{P}_i^{agg}$ as shown in (15.b). That is, we condense the massive state variables of ACs within node $i$ into just two sets of aggregated state variables, as shown in (25). This condensation enables us to efficiently fit the value function. Accordingly, the decision variables following state variable aggregation can be reformulated as shown in (26).

$$\boldsymbol{S}_t^{agg} = \left\{ T_{i,t}^{a,agg}, T_{i,t}^{m,agg} \right\} \tag{25}$$

$$\boldsymbol{X}_t^{agg} = \left\{ \begin{array}{l} P_{i,t}^{buy}, P_{i,t}^{sell}, P_{i,t}^{eg}, Q_{i,t}^{eg}, P_{i,t}^{dg}, Q_{i,t}^{dg}, P_{i,t}^{dru}, P_{i,t}^{drd}, \\ P_{i,t}^{agg}, Q_{i,t}^{agg}, P_{i,t}^{wd}, Q_{i,t}^{wd}, P_{i,t}^{pv}, Q_{i,t}^{pv}, V_{j,t}, \theta_{j,t}, P_{ij,t}^{L} \end{array} \right\} \tag{26}$$

Due to the convexity of the scheduling model (1.a)-(1.h) and the aggregate set $\mathbb{P}_i^{agg}$, the post-decision value function $V_t^x$ has a convex function relationship with aggregate states $\boldsymbol{S}_t^{agg}$ [30]. Therefore, a set of convex piece-wise linear functions can be applied for effective value function fitting around $\boldsymbol{S}_t^{agg}$, where the slopes are monotonically increasing to ensure convexity [29]. The advantages of using convex piecewise linear functions for value function fitting include high fitting accuracy and low computational burden. The convex piece-wise linear functions are illustrated in Fig. 4 and formulated as shown in (27.a)-(27.c).

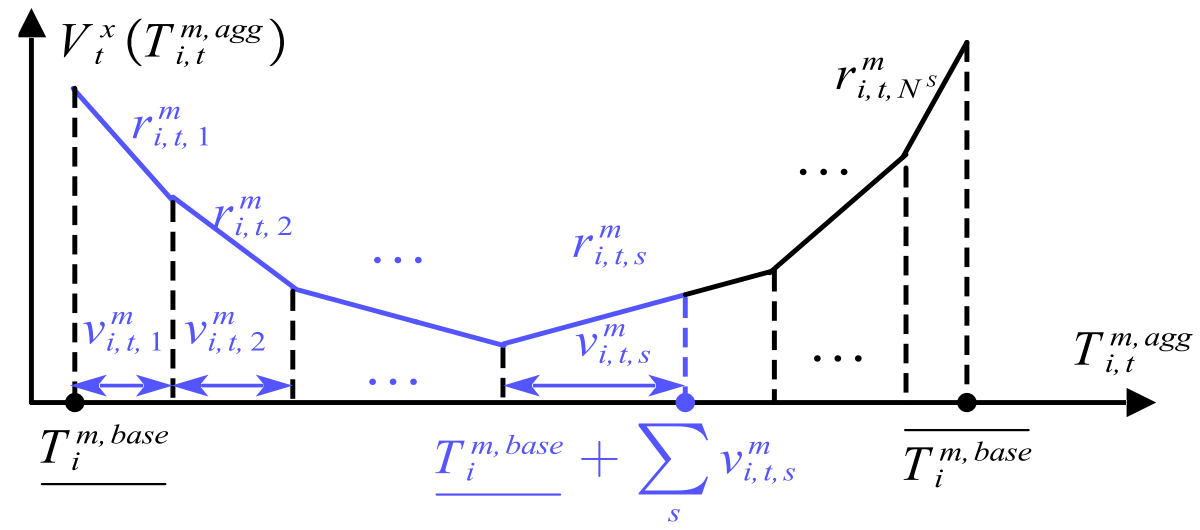

Fig. 4. Illustration of the value function construction via convex piece-wise linear functions.

$$V_t^x\left(T_{i,t}^{a,agg}, T_{i,t}^{m,agg}\right) = V_t^x\left(T_{i,t}^{a,agg}\right) + V_t^x\left(T_{i,t}^{m,agg}\right)$$
$$= \sum_i \sum_s (r_{i,t,s}^a v_{i,t,s}^a) + \sum_i \sum_s (r_{i,t,s}^m v_{i,t,s}^m) \tag{27.a}$$

$$0 \le v_{i,t,s}^m \le \left(\overline{T_i^{m,base}} - \underline{T_i^{m,base}}\right) / N^S, 0 \le v_{i,t,s}^a \le 2\beta_i^{set,base} / N^S \tag{27.b}$$

$$\underline{T_i^{m,base}} + \sum_s v_{i,t,s}^m = T_i^{m,agg}, \quad T_i^{set,base} - \beta_i^{set,base} + \sum_s v_{i,t,s}^a = T_i^{a,agg} \tag{27.c}$$

(27.a) represents the value function construction via convex piece-wise linear functions; $s$ and $N^S$ are index and the total number of the piece-wise segments (i.e., each aggregated state variable is divided into $N^S$ segments); $r^a/r^m$ represents the segment slope of the segment $s$ for $T^{a,agg}$ /$T^{m,agg}$, which is a parameter that need to be carefully predetermined; $v^a$ / $v^m$ captures the temperature quantity allocated to segment $s$, which is a variable that need to be optimized.

(27.b) represents the bounds of $v^a$/$v^m$ since each aggregated state variable is evenly divided into $N^S$ segments; $\underline{T^{m,base}}$ /$\overline{T^{m,base}}$ is the minimum/maximum parameter of state variable $T^{m,agg}$ in $\mathbb{U}_i^{base}$, i.e., $\underline{T_i^{m,base}} \le T_i^{m,agg} \le \overline{T_i^{m,base}}$; $\beta^{set,base}$/$T^{set,base}$ indicates the temperature set-point/tolerance parameter of state variable $T^{a,agg}$ in $\mathbb{U}_i^{base}$, i.e., $T_i^{set,base} - \beta_i^{set,base} \le T_i^{a,agg} \le T_i^{set,base} + \beta_i^{set,base}$.

(27.c) shows that the post-decision state can be calculated by accumulating quantities of all segments; With the convexity property of value functions and scheduling model, no value can be assigned to later segments until former ones are fully filled up [16].

After the values of segment slopes, i.e., $r^a$ and $r^m$, are carefully selected, we can treat $\boldsymbol{S}_{t+1}^{agg}$, $\boldsymbol{X}_t^{agg}$, $v^a$, and $v^m$ as decision variables to recast the original Bellman equation (8) as a linear program (28). Accordingly, the near-optimal value of decision variables $\boldsymbol{X}_t^{agg,*}$ in period $t$ can be obtained via solving (28).

$$V_t(\boldsymbol{S}_t^{agg}) = \min \left\{ \begin{array}{l} C_t^{obj}\left(\boldsymbol{S}_t^{agg}, \boldsymbol{X}_t^{agg}\right) + \\ \sum_i \sum_s (r_{i,t,s}^a v_{i,t,s}^a) + \sum_i \sum_s (r_{i,t,s}^m v_{i,t,s}^m) \end{array} \right\} \tag{28}$$

s.t. (1.b),(15.b),(1.d)-(1.h),(27.b)-(27.c)

Next, to ensure the quality of value function fitting, we need to properly select the segment slopes $r^a$ and $r^m$. Herein, we employ the sample slope estimation algorithm [34], known for its favorable convergence, to train the slopes of convex piece-wise linear functions offline using historical data, as outlined in **Algorithm 1**. In the $\kappa th$ iteration, the sample estimation of the marginal value $C_{i,t,\kappa}^{[\cdot],\delta}$ can be calculated via (29). Then, the estimated slopes $r_{i,t-1,s,\kappa}^{[\cdot]}$ can be updated via (30).

$$\begin{cases} C_{i,t,\kappa}^{a,\delta}\left(T_{i,t,\kappa}^{a,agg}\right) = \partial V_{t,\kappa}(\boldsymbol{S}_t^{agg}) / \partial T_{i,t,\kappa}^{a,agg} \\ C_{i,t,\kappa}^{m,\delta}\left(T_{i,t,\kappa}^{m,agg}\right) = \partial V_{t,\kappa}(\boldsymbol{S}_t^{agg}) / \partial T_{i,t,\kappa}^{m,agg} \end{cases} \tag{29}$$

$$r_{i,t-1,s^*,\kappa}^{[\cdot]} = (1-\xi) r_{i,t-1,s^*,\kappa-1}^{[\cdot]} + \xi C_{i,t,\kappa}^{[\cdot],\delta}$$

$$r_{i,t-1,s,\kappa}^{[\cdot]} = \begin{cases} \max\left\{ r_{i,t-1,s,\kappa-1}^{[\cdot]}, r_{i,t-1,s^*,\kappa}^{[\cdot]} \right\}, s > s^* \\ r_{i,t-1,s^*,\kappa}^{[\cdot]}, s = s^* \\ \min\left\{ r_{i,t-1,s,\kappa-1}^{[\cdot]}, r_{i,t-1,s^*,\kappa}^{[\cdot]} \right\}, s < s^* \end{cases} \tag{30}$$

where $s^*$ is the optimized segment index derived from (29) and $\xi$ is the step size parameter for updating slopes. Equation (30) is based on the leveling algorithm [35], which helps maintain the convexity of the value function.

**Algorithm 1** Sample slope estimation

**Initialize**: Set iterations $\kappa = 1$, initial slopes $r_{i,t,s,\kappa}^{[\cdot]}$, and parameter $\xi$.
**Procedure:**
**Step 1:** Use Monte Carlo method to generate a training scenario.
**Step 2:** Offline train the slopes of value function via the generated scenario.
**For** $t = 1, \ldots, N^T$:
  Observe the uncertainty realization in period $t$;
  Determine the optimal decision variables in period $t$ via (28);
  Update the slopes via (29)-(30);
  Update the system state in period $t$ via state transition.
**End for**
**Step 3:** Let $\kappa = \kappa + 1$. If $\kappa \le N^\kappa$, go to Step 1;

*E. Practical Application Framework*

The practical application framework of the proposed TCA-ADP scheduling algorithm consists of the following three sequential processes:

***Process 1** (Offline State Variable Aggregation)*: The time-causal state variable aggregation is conducted via calculating (14.a)-(14.c) to derive the aggregate set $\mathbb{P}_i^{agg}$, as described in (15.b). Note that in this process, due to the orthogonality among the aggregation parameters of different ACs, parallel computing techniques can be employed to calculate the aggregation parameters for massive ACs in (24.a)-(24.b). This parallel-enabled feature provides significant scalability to the proposed state variable aggregation model.

***Process 2** (Offline Value Function Fitting)*: The value functions of aggregated state variables are fitted using historical data via **Algorithm 1**.

***Process 3** (Online Real-time Application):* See **Algorithm 2**.

**Algorithm 2** Online real-time application

**Initialize**: Set initial period $t = 1$.
**Procedure:**
**Step 1:** Observe the information in period $t$.
**Step 2**: Decide the optimal decision variables in period $t$ by (28) via the well-trained value functions, and update the current system state via state transition.
**Step 3**: Disaggregate the decision results to each AC.
**Step 4:** Let $t = t + 1$. If $t \le N^T$, go to Step 1.

In **Algorithm 2 Step 3**, when the optimal value of aggregate power profile $P_{i,t}^{agg,*}$ is known after solving the real-time scheduling problem in period $t$, there is no need to solve an additional optimization problem to allocate the power profile for each individual AC. Instead, the disaggregated power profiles $P_{i,k,t}^{a,*}$ for each AC can be directly expressed as explicit functions of the aggregate power profile $P_{i,t}^{agg,*}$ as shown in (31). The derivation of (31) is presented in Appendix D.

$$P_{i,k,t}^{a,*} = \gamma_{i,k,t}^{aff} + \Gamma_{i,k,t}^{aff}\left(\sum_k(\Gamma_{i,k,t}^{aff})\right)^{-1}\left(P_{i,t}^{agg,*} - \sum_k(\gamma_{i,k,t}^{aff})\right) \quad (31)$$

Note that this disaggregation calculation is both parallelizable and closed-form, ensuring scalable, fast, and feasible power allocation for massive ACs. Specifically, the use of inner approximation in state variable aggregation guarantees feasibility, while parallel and closed-form computation supports both speed and scalability.

## IV. Case Studies

Case studies on the modified 8-bus and IEEE 123-bus test systems are conducted to validate the effectiveness and scalability of the proposed TCA-ADP algorithm. The numerical simulations, programmed in MATLAB, are solved using GUROBI 11.0.0 on a 2.30 GHz CPU with 16 GB RAM.

First, the 8-bus system is analyzed using data from [36]. The scheduling horizon spans 24 hours with 1-hour granularity. One deterministic forecast (represented by a line with dots) and multiple uncertain scenarios are depicted in Fig. 5, generated from historical data for value function training and real-time testing [37]. There are 40 ACs within bus 8. The profiles of outdoor temperature and aggregate reference power for ACs are illustrated in Fig. 6. The heterogeneous AC parameters are outlined in TABLE II. The cost prices $c^{gt}$ and $c^{dr}$ are set as 0.5 \$/kWh and 0.005 \$/kWh [29].

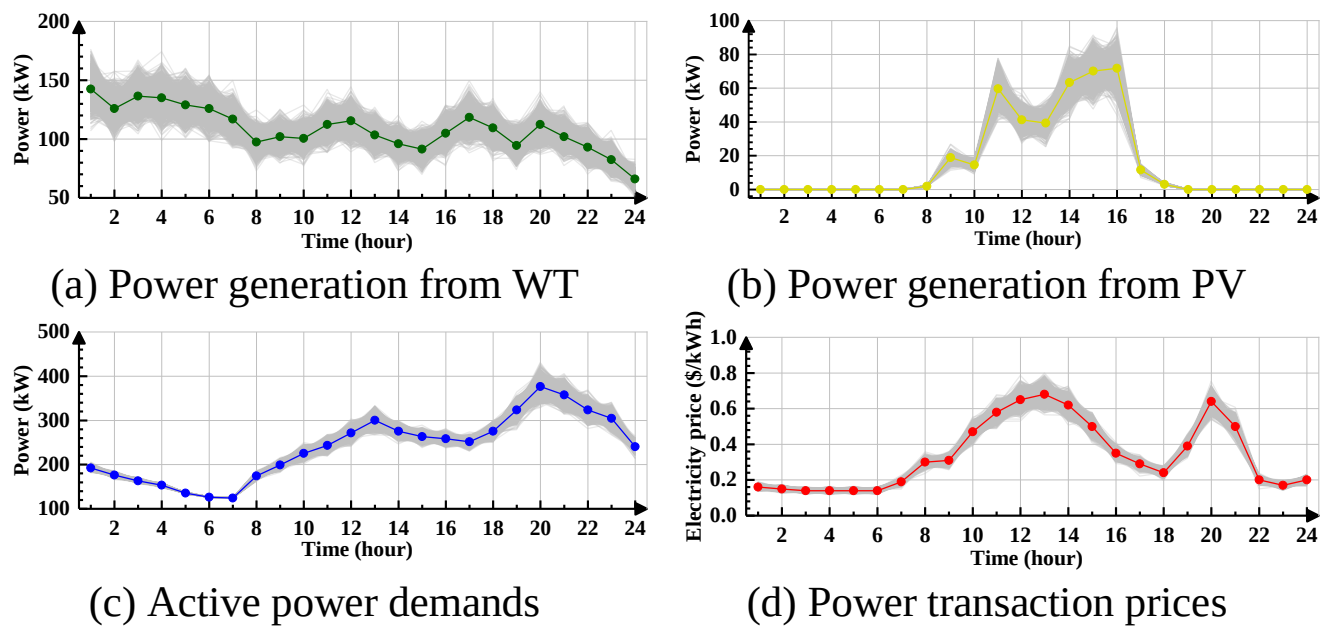

(a) Power generation from WT (b) Power generation from PV

(c) Active power demands (d) Power transaction prices

Fig. 5. Deterministic forecast and uncertainty scenarios.

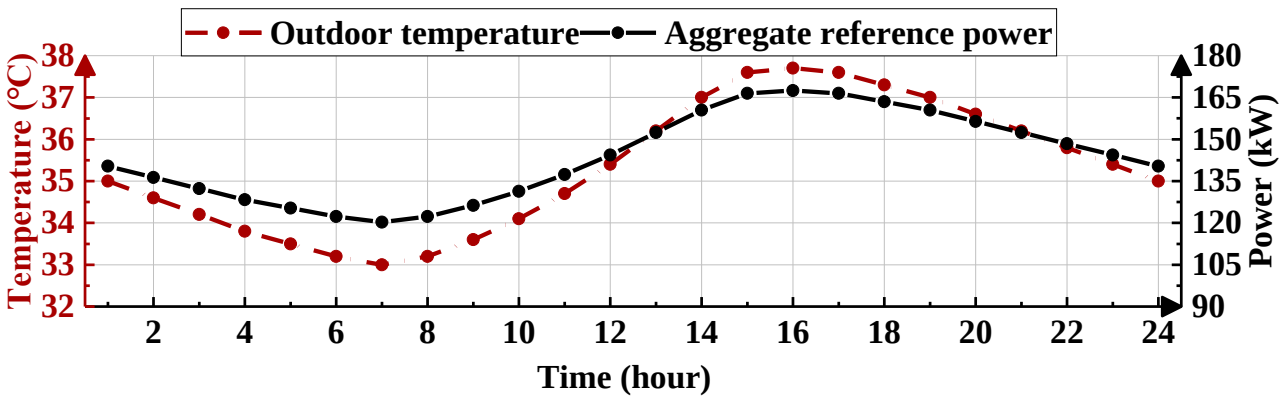

Fig. 6. Outdoor temperature and reference power for ACs.

TABLE II
AC Parameters in 8-Bus System

| Parameter | Value | Unit | Parameter | Value | Unit |
|---|---|---|---|---|---|
| $H^a$ | 0.15-0.35 | °C/kW | $H^m$ | 0.2-0.8 | °C/kW |
| $C^a$ | 0.5-3.5 | kWh/°C | $C^m$ | 4-8 | kWh/°C |
| $T^{set}$ | 20-22 | °C | $\beta^{set}$ | 1-3 | °C |
| $\mu^h$ | 0.95 | p.u. | $f^h$ | 0.1 | p.u. |
| $\underline{P^a}$ | 1.5 | kW | $\overline{P^a}$ | 6.5 | kW |

*A. Validation of TCA-ADP Algorithm*

We begin by comparing the following cases in the deterministic scenario to validate the TCA-ADP algorithm:

**Case 0 (Benchmark)**: Ideal situation algorithm based on perfect prediction without aggregation, providing hindsight and theoretically optimal results in one shot. However, it violates time causality and is impossible in reality, thereby only serving as a lower bound for evaluation. We refer to the *solution gap* as the absolute relative deviation of the total cost between **Benchmark** and other cases.

**Case 1 (Proposed)**: TCA-ADP algorithm in this paper.

**Case 1a**: ADP algorithm where the uniform-scaling method from [32] is used for state variable aggregation.

**Case 1b**: ADP algorithm where the state variables of massive ACs are considered without aggregation.

*1) Real-time Scheduling Results of ADN:* The real-time scheduling results are reported in TABLE III and Fig. 7-Fig. 9.

TABLE III
Scheduling Results in 8-Bus System After Training

| Item | Case 0 | Case 1 | Case 1a | Case 1b |
|---|---|---|---|---|
| Total cost (\$) | 659.5 | 679.5 | 773.5 | 664.5 |
| Solution gap (%) | / | 3.03 | 17.29 | 0.76 |
| AC response energy (kWh) | 1,557.2 | 1,448.7 | 1,112.5 | 1,520.1 |
| Average CPU time in one training iteration (s) | / | 0.61 | 0.58 | 182.45 |

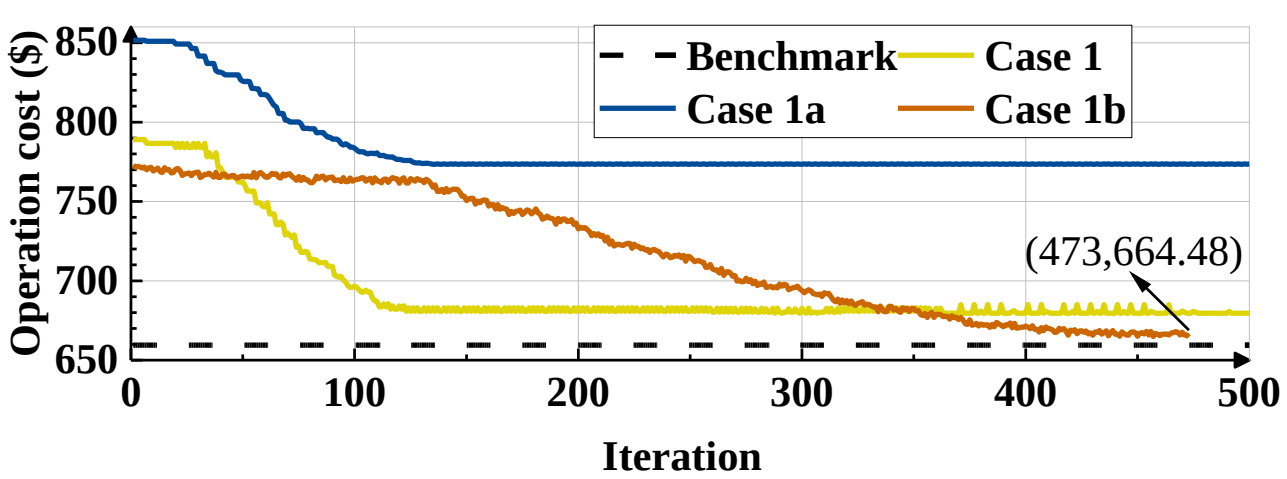

Fig. 7. Training profiles for different cases.

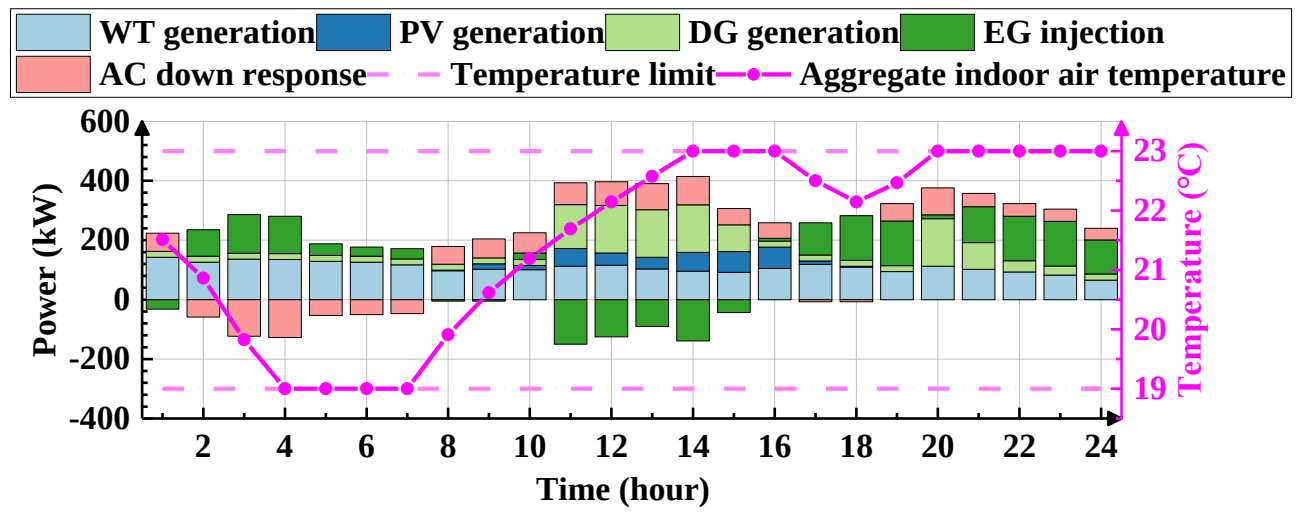

Fig. 8. ADN scheduling results for **Case 1**.

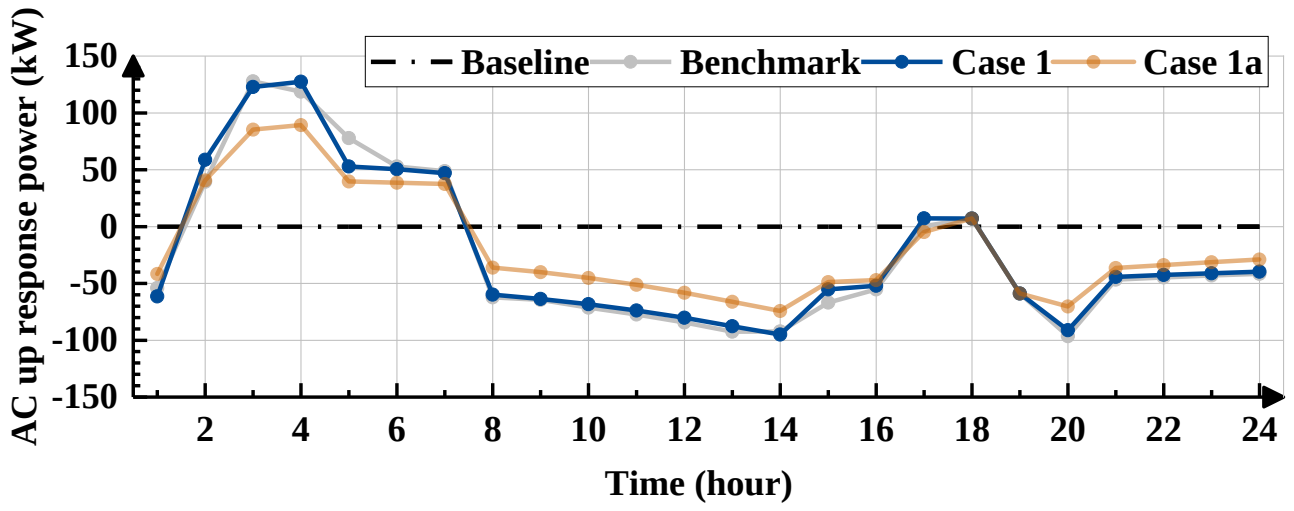

Fig. 9. Aggregate response power profiles for ACs.

The training profiles for different cases are shown in Fig. 7. From TABLE III and Fig. 7, we observe that although **Case 1b** shows the lowest solution gap of 0.76% after 473 iterations, it requires significantly more CPU time during training process, about 182.45 seconds per iteration. Consequently, only 473 iterations can be conducted within an offline period of 24 hours. Additionally, **Case 1b** suffers slow convergence speed due to the need to update massive value function slopes simultaneously, which limits its applicability for real-world scheduling of massive ACs. In contrast, both **Case 1** and **Case 1a** benefit from state variable aggregation and quickly converge after 500 iterations, with each iteration taking only a few seconds. Among these, **Case 1** has a relatively low solution gap of 3.03%. This validates the convergence performance of the proposed TCA-ADP algorithm in the offline training process.

The ADN scheduling results for **Case 1** are given in Fig. 8. From Fig. 8, we observe that the power response characteristics of ACs are closely linked to electricity prices. Specifically, during periods of low electricity prices, the ACs perform an up response by increasing their electricity demand to store energy as heat. This is reflected in the decrease in the aggregate indoor temperature, observed during periods such as 2-7 and 17-18. In contrast, during periods of high electricity prices, the ACs perform a down response, releasing the stored thermal energy to reduce electricity demand, leading to an increase in the aggregate indoor temperature, as seen in periods 8-16 and 19-24. In this way, ACs can leverage temporal differences in electricity prices, collaborating with other generators to shift power loads and reduce electricity demand during high-price periods. This indicates that the proposed TCA-ADP algorithm effectively leverages the AC response characteristics to coordinate power generation and electricity transactions, thereby enhancing the overall economic efficiency of the ADN.

The response characteristics observed above are further validated in Fig. 9. From Fig. 9, we observe that the AC response characteristics are consistent across all three cases. In particular, **Case 1** demonstrates more comparable scheduling results to **Benchmark** than **Case 1a**. This is also confirmed by TABLE III, where **Case 1** identifies 93.09% of the cost-effective response energy in **Benchmark**. This result is attributed to the affine transformation proposed in **Case 1**, which provides a more flexible transformation approach, thereby improving the accuracy of state variable aggregation. In contrast, **Case 1a** identifies only 71.44% of the cost-effective response energy in **Benchmark**. This is because **Case 1a** limits the inner approximation structure, sacrificing much aggregate flexibility of the ACs and, as a result, reducing their response capacity. This validates that the proposed state variable aggregation achieves a more satisfactory aggregation accuracy and effectively captures the aggregate flexibility.

*2) Real-time Disaggregation Results of massive ACs*: The online disaggregation results of ACs in **Case 1** are given in Fig. 10 and Fig. 11.

From the figures, we observe that the operational constraints of each AC, in terms of both power and indoor air temperature limits, are strictly satisfied due to the inner approximation of the proposed state variable aggregation. This indicates that the proposed TCA-ADP algorithm efficiently and feasibly disaggregates the aggregated AC power and allocates it to each individual AC, ensuring the practical applicability of the TCA-ADP algorithm.

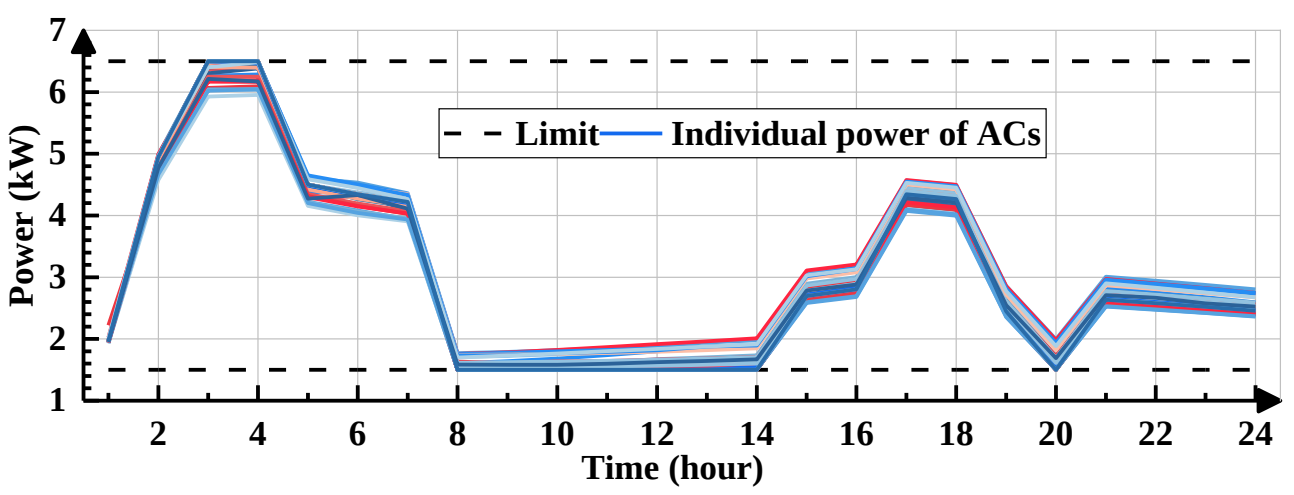

Fig. 10. Disaggregated power profiles in **Case 1**.

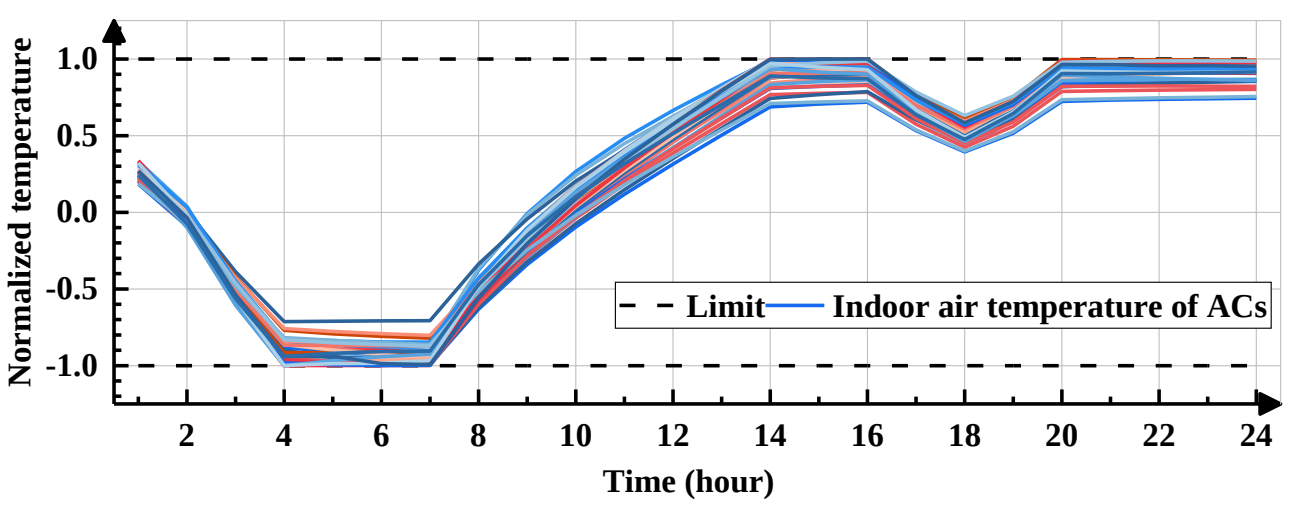

Fig. 11. Disaggregated temperature profiles in **Case 1**.

**Normalized temperature* is defined as $(T^a - T^{set})/\beta^{set}$ for analysis and bounded between -1 and 1 according to (1.c).

*3) Sensitivity Analysis*: We further conduct sensitivity analysis on the price of AC response, i.e., $c_i^{dr}$. The results are

given in Fig. 12.

From Fig. 12, we observe that as the price for leveraging AC flexibility $c_i^{dr}$ increases, the ADN's demand for AC response energy decreases, while its real-time scheduling cost rises. When $c_i^{dr}$ reaches 1 \$/kWh, using AC flexibility becomes too expensive, and thus, the ADN no longer utilizes AC response energy. This suggests that $c_i^{dr}$ plays a key role in determining the level of AC flexibility utilization. A lower $c_i^{dr}$ makes AC flexibility more cost-effective, encouraging the ADN to use it more, but setting it too low could reduce user participation. Therefore, determining an appropriate $c_i^{dr}$ is crucial to balance ADN scheduling costs and user engagement.

Additionally, the proposed state variable aggregation method consistently identifies more response energy than the uniform-scaling method, leading to near-optimal scheduling outcomes, close to **Benchmark**.

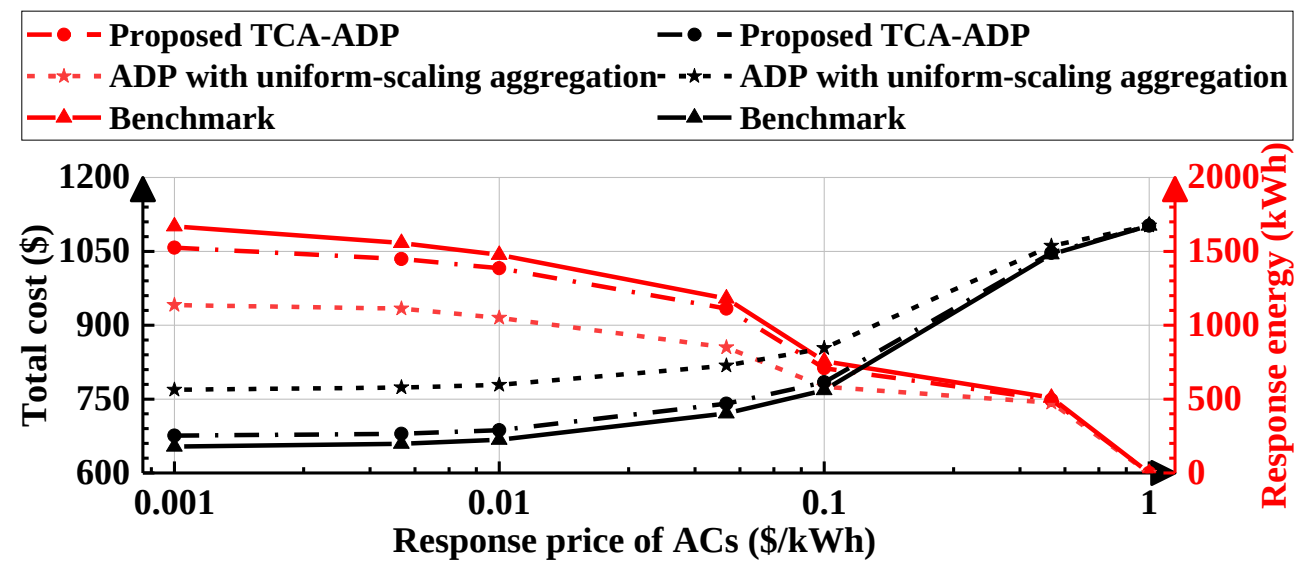

Fig. 12. Sensitivity analysis on the price of AC response.

In summary, the above results show that the proposed TCA-ADP algorithm has favorable aggregation accuracy for massive state variables of ACs, thereby providing comparable economic scheduling results to the ideal situation.

*B. Comparison with Other Real-time Scheduling Algorithms*

The proposed TCA-ADP algorithm is further compared with other real-time scheduling algorithms to show its advantages.

**Case 2**: Myopic algorithm where the scheduling results in period t are obtained by solving a single-period optimization problem involving only information and variables in period $t$.

**Case 3&4**: MPC algorithm, where the scheduling results in period $t$ are obtained via solving a H-period look-ahead optimization problem, utilizes both current observed information and predicted future information (from period $t$ to future period $t + H$) to optimize the decision-making process (i.e., MPC-H). In **Case 3**, we set $H = 4$, and in **Case 4,** $H = 8$, based on the requirements for forecasting accuracy.

*1) Deterministic Scenario:* We first use the deterministic scenario for comparison. The scheduling results are reported in TABLE IV and Fig. 13.

TABLE IV
SCHEDULING COMPARISON IN DETERMINISTIC 8-BUS SYSTEM

| Item | Case 1 | Case 2 | Case 3 | Case 4 |
|---|---|---|---|---|
| Total cost ($) | 679.5 | 788.9 | 759.2 | 726.3 |
| Solution gap (%) | 3.03 | 19.62 | 15.12 | 10.13 |

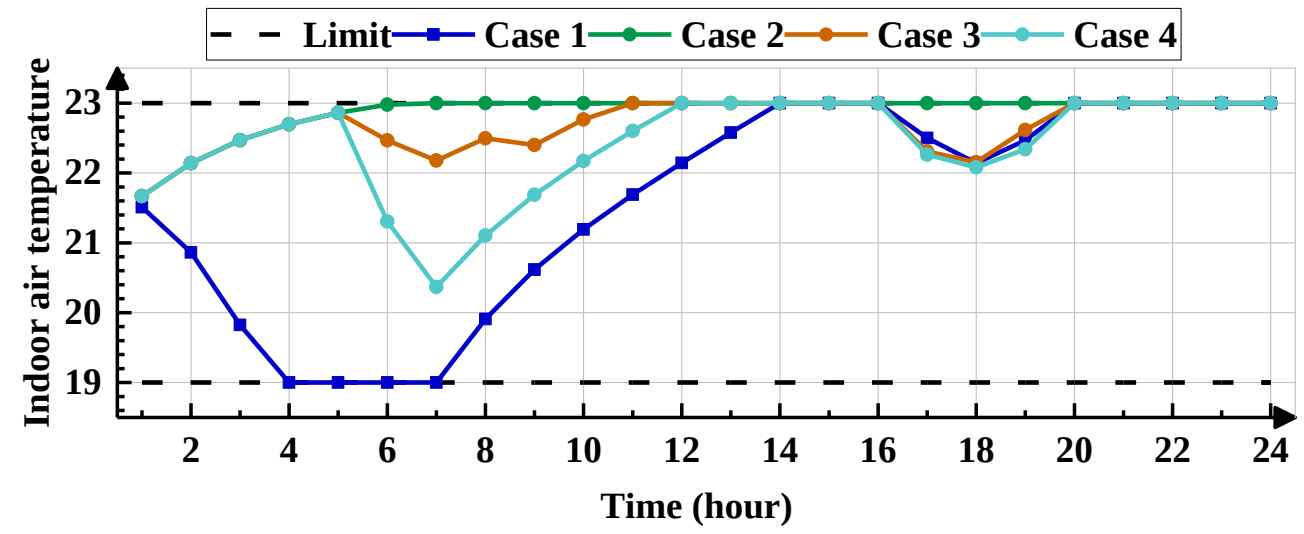

Fig. 13. Indoor air temperature profile for different algorithms.

From TABLE IV, we observe that both the Myopic and MPC algorithms yield local optimal solutions with relatively high solution gaps of 19.62%, 15.12%, and 10.13% in **Cases 2–4**, respectively. In contrast, the proposed TCA-ADP algorithm in **Case 1** achieves a global near-optimal solution across the entire horizon with the lowest gap of 3.03%.

This result is further validated by the analysis in Fig. 13, which shows the profiles of indoor air temperature for different algorithms. From Fig. 13, we observe that the scheduling results in **Case 2-4** differ markedly from those in **Case 1**. This is because both Myopic and MPC algorithms rely solely on current or near-future forecast information, which is relatively shortsighted.

Specifically, the AC response decisions in **Case 2** are not sensitive to electricity price variations; for example, during the lower-price period 2–7, ACs release all stored thermal energy to reduce electricity demand, leading to a situation where, in the higher-price period 8–14, the indoor temperature reaches its upper limit and there is little room for adjustment. As a result, **Case 2** shows the poorest economic performance.

In addition, the MPC algorithm in **Case 3-4** incorporates near-future information into current decisions, allowing it to respond to near-future price signals. As a result, its scheduling outcome is closer to the one in **Case 1**. For instance, in **Case 4**, during the lower-price period 5–7, MPC anticipates the price increase in periods 8–16 and stores thermal energy in advance, thus improving scheduling economics. Furthermore, we observe that the solution gap in **Case 4** is lower than in **Case 3**, as MPC-8 takes into account more future periods. However, the MPC algorithm is highly sensitive to the accuracy of forecasted future information, and its performance can degrade significantly if predictions are inaccurate. Moreover, when future information is unavailable, the MPC algorithm reduces to the Myopic one.

In contrast, the proposed TCA-ADP algorithm addresses this issue by embedding uncertainty information into the slope of the trained value function during the offline training process, enabling it to perform global and near-optimal scheduling in real-time without relying on exact future predictions.

*2) Uncertain Scenarios:* We further conduct stochastic simulations under multiple uncertain scenarios for comparison (using data in Fig. 5). There are 1,200 scenarios for offline training and 200 scenarios for real-time testing. The scheduling results are shown in Fig. 14-Fig. 15.

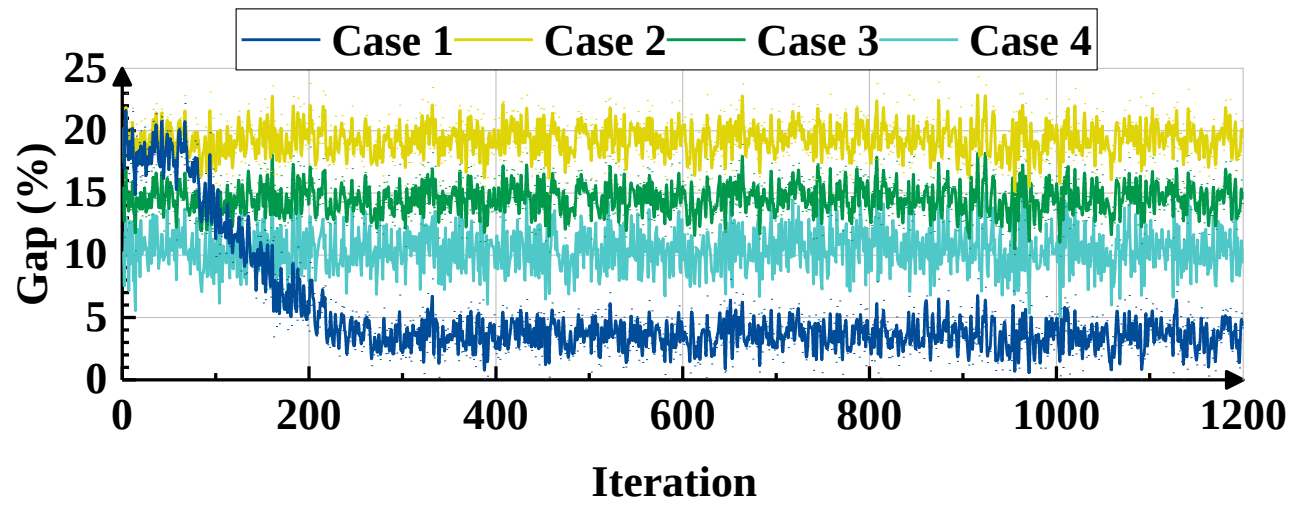

Fig. 14. Training profile in 8-bus System.

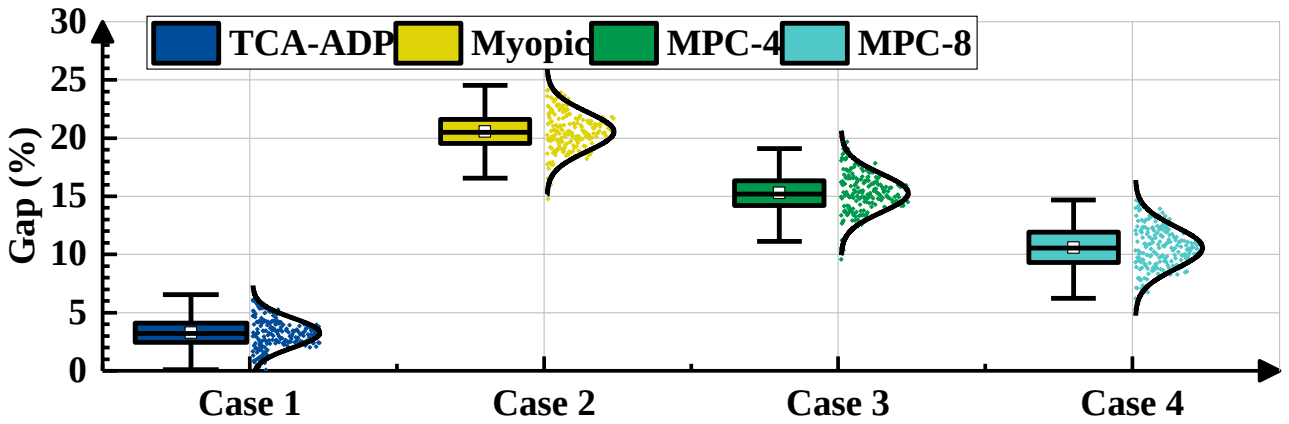

Fig. 15. Real-time testing results in 8-bus System.

From Fig. 14, we observe that the solution gap of **Case 1** starts from a relatively high value close to **Case 2**, and gradually approaches around 3% via updating value function slopes from historical data. After 1,200 iterations, the solution gap of **Case 1** becomes the lowest among all cases. It indicates that the proposed TCA-ADP algorithm contributes to learning from the historical data and converging to the global near-optimal solutions under time-causal uncertainties. On the other hand, the solution gaps of **Case 2-4** do not show a convergence trend in the offline training, remaining high throughout the iterations.

From Fig. 15, we observe that, after informative training and learning, the well-trained value functions guide the ACs towards near-optimal operating conditions. As a result, the average solution gap in **Case 1** is approximately 3.5% across 200 real-time testing scenarios, demonstrating much better performance compared to other cases.

In a nutshell, the above results verify that the proposed TCA-ADP algorithm effectively offline trains the value functions of massive ACs using historical data, thus fully exploiting the ACs' flexibility from a global perspective to address time-causal uncertainties in practical online applications.

### *C. Scalability Test*

We further verify the computation tractability and scalability of the proposed TCA-ADP algorithm using the IEEE 123-bus system including 12 DGs, 13 WTs, 8 PVs, 1 EG, and 1,000 ACs distributed across three groups, with 300, 400, and 300 units in each group, respectively (parameters from TABLE V). The electricity price, and per-unit output values of the load, WT and PV are the same as that in the modified 8-bus system. The rated rigid load is 6,326 kW. The WT and PV capacity are scaled up to 6,000 kW and 1,400 kW respectively. Other detailed parameters can be assessed in [36]. Training and testing scenarios are set at 1,200 and 200, respectively.

*1) Real-time Scheduling Results*: The real-time scheduling results are reported in Fig. 16 and TABLE VI.

The training profiles are given in Fig. 16. We observe that in larger test system, the proposed TCA-ADP performs best with fast convergence in value function training and merely around 4% solution gap after 1,200 iterations. However, the traditional ADP algorithm without aggregation in **Case 1b** no longer converges in the larger-size system. In **Case 1b**, one iteration takes about 9,000 s CPU time for value function training in **Algorithm 1 Step 2**. Indeed, for 1,000 ACs, 2*1,000*2*24= 96,000 optimization problems must be solved per iteration. Here, 2*1,000 represents the number of state variables per period, the second factor of 2 corresponds to the number of slope samples (left and right), and 24 denotes the number of periods. Thus, value functions can only be updated in 9 iterations in **Case 1b** each day which are too few for value function training. In contrast, one iteration takes only about 6.8 s in **Case 1** where only 2*3*2*24=288 optimization problems need solved per iteration. Also, **Case 1** benefits from much smaller optimization scale and faster convergence speed due to state variable aggregation. Thus, the proposed TCA-ADP algorithm is necessary for solving realistic operation problems.

The real-time testing results are given in TABLE VI. We observe that **Case 1** still has the smallest testing solution gap due to the well-approximate value functions trained from historical data. Thanks to the state variable aggregation, **Case 1** also shows decent computational efficiency with only average 0.52 s CPU time for each scenario in online applications.

TABLE V
AC PARAMETERS IN 123-BUS SYSTEM

| Parameter | Group 1 | Group 2 | Group 3 |
|---|---|---|---|
| $H^a$ (°C/kW) | 0.1-0.25 | 0.1-0.15 | 0.25-0.4 |
| $H^m$ (°C/kW) | 0.5-2 | 0.3-1 | 0.1-0.2 |
| $C^a$ (kWh/°C) | 1-3 | 2.5-5.5 | 3-7 |
| $C^m$ (kWh/°C) | 3-6 | 3-5 | 5-8 |
| $T^{set}$ (°C) | 17-21 | 19-23 | 20-22 |
| $\beta^{set}$ (°C) | 1-3 | 1-2 | 2-3 |
| $\mu^h$ | 0.9-0.95 | 0.92-0.94 | 0.92-0.95 |
| $f^h$ | 0.08-0.1 | 0.05-0.1 | 0.08-0.12 |
| $\underline{P^a}$ (kW) | 0.5-1 | 0.3-0.9 | 0.5-2.5 |
| $\overline{P^a}$ (kW) | 6-7 | 4-6 | 7.5-8.5 |

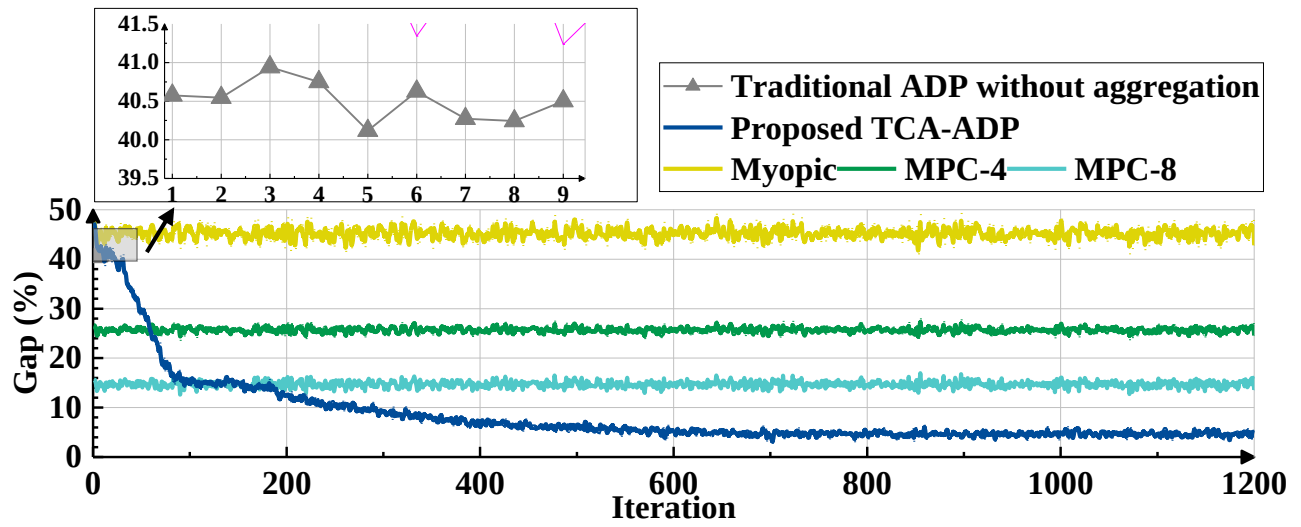

Fig. 16. Training profile in 123-bus System.

TABLE VI
REAL-TIME TESTING RESULTS IN 123-BUS SYSTEM

| Item | **Case 1** | **Case 2** | **Case 3** | **Case 4** |
|---|---|---|---|---|
| Average total cost in online scenario testing (k$) | 10.69 | 14.91 | 12.89 | 11.65 |
| Average solution gap in online scenario testing (%) | 4.08 | 45.18 | 25.51 | 13.44 |
| Average CPU time in online scenario testing (s) | 0.52 | 5.62 | 25.56 | 104.99 |

*2) Sensitivity Analysis:* We further conduct sensitivity analysis on the number of ACs in value function training and

the results after up to one day of iteration are given in Fig. 17.

From Fig. 17, we observe that the iteration time remains nearly constant in **Case 1**, even as the scale of ACs increases. This stability is due to the proposed aggregation method, which allows each AC to be independently and parallelly approximated, as described in Section III-E. In contrast, the iteration time of **Case 1b** significantly increases with scale. When the number of ACs exceeds 100, traditional ADP algorithm without aggregation fail to converge within one day of training. Due to the insufficient number of iterations, the value function of **Case 1b** cannot be well approximated, which prevents the solution gap from being reduced. As a result, the solution gap of **Case 1b** eventually approaches that of the shortsighted Myopic method, leading to uneconomic scheduling. In contrast, **Case 1** has a relatively stable solution gap of around 4% because the effectiveness of the proposed state variable aggregation method is not significantly affected by the scale of ACs.

The above results show that the proposed TCA-ADP algorithm balances economy and tractability against the large value function space of massive ACs, thereby exhibiting high scalability for practical implementation.

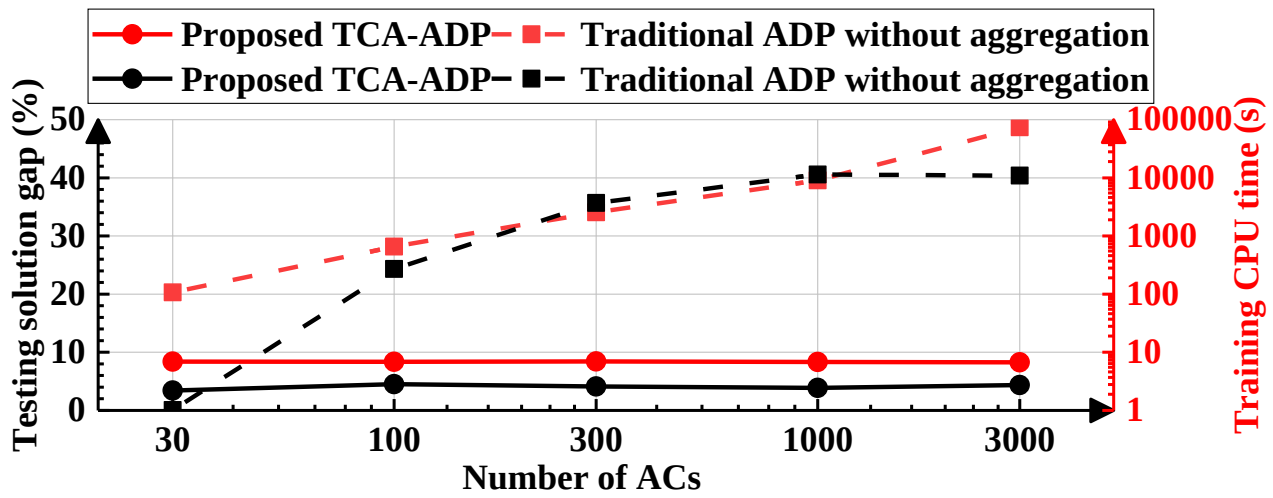

**Testing solution gap* refers to average solution gap in testing scenarios after training within an offline period of 24 hours and *Training CPU time* refers to average CPU time in one scenario during offline training.

Fig. 17. Sensitivity analysis on the number of ACs.

## V. Conclusions

This paper focuses on leveraging time-causal state variable aggregation for real-time schedule of massive ACs under renewable uncertainty. The time-causality requirement for state variable aggregation in real-time scheduling is discussed. Given this, a novel TCA-ADP algorithm is introduced to efficiently schedule massive ACs. Through case studies, we observe that the proposed TCA-ADP algorithm 1) shows favorable aggregation accuracy for near-optimal economic scheduling; 2) globally exploits ACs' flexibility to address time-causal uncertainties in online applications; 3) balances economy and tractability against the large value function space of massive ACs for practical implementation.

While the traditional ADP algorithm without aggregation performs well in small-scale systems with few ACs, it struggles with convergence in large-scale systems with massive ACs due to heavy computational burden. In contrast, the proposed TCA-ADP algorithm, while inevitably introducing a solution gap of around 4%, achieves significant scalability by condensing the value function space through state variable aggregation, enabling practical real-time scheduling of massive ACs.

Future research directions include the following. First, the proposed state variable aggregation model will be extended to incorporate the impact of real-time uncertainties, such as outdoor temperature, in time-causal aggregation. Second, an exploration of the model's application to nonconvex flexible loads, such as heat pumps and water heaters with binary variables and non-connected feasible regions, is planned. Lastly, integrating additional practical market factors into the TCA-ADP algorithm, including time-of-use demand charges [38] and fair distribution of disaggregated AC power, would be an interesting advancement.

## Appendix

### A. Coefficient Matrices and Vectors in Illustrative Case

$$\boldsymbol{H}=\begin{bmatrix}1.86 & 0 & 0\\ -1.17 & 1.86 & 0\\ 0 & -1.17 & 1.86\\ -1.86 & 0 & 0\\ 1.17 & -1.86 & 0\\ 0 & 1.17 & -1.86\\ 0 & -1 & 0\\ 0 & 0 & -1\\ 0 & 1 & 0\\ 0 & 0 & 1\end{bmatrix},\ \boldsymbol{h}=\begin{bmatrix}41.45\\ 16.73\\ 15.51\\ -35.45\\ 10.73\\ 9.51\\ -19\\ -19\\ 23\\ 23\end{bmatrix}$$

$$\boldsymbol{H}^{base}=\begin{bmatrix}81.33 & 0 & 0\\ -51.25 & 14.66 & 0\\ 0 & -9.24 & 27.65\\ -81.33 & 0 & 0\\ 51.25 & -14.66 & 0\\ 0 & 9.24 & -27.65\\ 0 & -7.88 & 0\\ 0 & 0 & -14.87\\ 0 & 7.88 & 0\\ 0 & 0 & 14.87\end{bmatrix},\ \boldsymbol{h}^{base}=\begin{bmatrix}1812.82\\ -804.03\\ 384.98\\ -1791.73\\ 826.42\\ -363.01\\ -167.46\\ -306.67\\ 180.32\\ 323.05\end{bmatrix}$$

### B. Removing Process for AC Sets

First, the thermal dynamic equations in (1.c) can be recast as below:

$$\begin{cases}\boldsymbol{A}_{i,k}^{c1}\boldsymbol{T}_{i,k}^{m}+\boldsymbol{A}_{i,k}^{c2}\boldsymbol{T}_{i,k}^{a}+\boldsymbol{A}_{i,k}^{c3}\boldsymbol{P}_{i,k}^{a}=\boldsymbol{A}_{i,k}^{c4}\\ \boldsymbol{A}_{i,k}^{c5}\boldsymbol{T}_{i,k}^{m}+\boldsymbol{A}_{i,k}^{c6}\boldsymbol{T}_{i,k}^{a}+\boldsymbol{A}_{i,k}^{c7}\boldsymbol{P}_{i,k}^{a}=\boldsymbol{A}_{i,k}^{c8}\end{cases} \tag{32.a}$$

where $\boldsymbol{A}_{i,k}^{[\cdot]}$ is the compact coefficient matrix. Since $\boldsymbol{A}_{i,k}^{c1}$, $\boldsymbol{A}_{i,k}^{c2}$, $\boldsymbol{A}_{i,k}^{c5}$, and $\boldsymbol{A}_{i,k}^{c6}$ are invertible matrix, we recast the constraint on the upper and lower limits of indoor temperature as follows.

$$T_{i,k}^{set}-\beta_{i,k}^{set}\le\boldsymbol{A}_{i,k}^{c10}-\boldsymbol{A}_{i,k}^{c9}\boldsymbol{P}_{i,k}^{a}\le T_{i,k}^{set}+\beta_{i,k}^{set} \tag{32.b}$$

where

$$\begin{cases}\boldsymbol{A}_{i,k}^{c9}=\boldsymbol{A}_{i,k}^{c11}\left(\boldsymbol{A}_{i,k}^{c7}-\boldsymbol{A}_{i,k}^{c5}\left(\boldsymbol{A}_{i,k}^{c1}\right)^{-1}\boldsymbol{A}_{i,k}^{c3}\right)\\ \boldsymbol{A}_{i,k}^{c10}=\boldsymbol{A}_{i,k}^{c11}\left(\boldsymbol{A}_{i,k}^{c8}-\boldsymbol{A}_{i,k}^{c5}\left(\boldsymbol{A}_{i,k}^{c1}\right)^{-1}\boldsymbol{A}_{i,k}^{c4}\right)\\ \boldsymbol{A}_{i,k}^{c11}=\left(\boldsymbol{A}_{i,k}^{c6}-\boldsymbol{A}_{i,k}^{c5}\left(\boldsymbol{A}_{i,k}^{c1}\right)^{-1}\boldsymbol{A}_{i,k}^{c2}\right)^{-1}\end{cases} \tag{32.c}$$

Thus, we obtain the equivalent flexibility set of the individual AC in power space as below:

$$\begin{cases}T_{i,k}^{set}-\beta_{i,k}^{set}\le\boldsymbol{A}_{i,k}^{c10}-\boldsymbol{A}_{i,k}^{c9}\boldsymbol{P}_{i,k}^{a}\le T_{i,k}^{set}+\beta_{i,k}^{set}\\ \underline{P_{i,k}^{a}}\le\boldsymbol{P}_{i,k}^{a}\le\overline{P_{i,k}^{a}}\end{cases} \tag{32.d}$$

Further, we can recast (32.d) as a compact H-representation form to derive the parameters $\boldsymbol{H}^{proj}$ and $\boldsymbol{h}^{proj}$ in (20).

*C. Proof of Containment Constraint Reformulation*

Given a base set as shown in (21), $\mathbb{U}_{i,k}^{aff} = \boldsymbol{\Gamma}_{i,k}^{aff}\mathbb{U}_i^{base} + \boldsymbol{\gamma}_{i,k}^{aff}$ can be expressed as:

$$\mathbb{U}_{i,k}^{aff} = \left\{ \boldsymbol{P}_{i,k}^{aff} = \boldsymbol{\Gamma}_{i,k}^{aff}\boldsymbol{P}_{i,k}^{a} + \boldsymbol{\gamma}_{i,k}^{aff}, \forall \boldsymbol{H}_i^{base,proj}\boldsymbol{P}_{i,k}^{a} \le \boldsymbol{h}_i^{base,proj} \right\} \tag{33.a}$$

where $\mathbb{U}_{i,k}^{aff}$ represents the affine transformed base set.

Also, $\mathbb{U}_{i,k}^{aff}$ can be expressed in a H-representation form:

$$\mathbb{U}_{i,k}^{aff} = \{ \boldsymbol{H}_{i,k}^{aff,proj}\boldsymbol{P}_{i,k}^{aff} \le \boldsymbol{h}_{i,k}^{aff,proj} \} \tag{33.b}$$

where $\boldsymbol{H}_{i,k}^{aff,proj}$/ $\boldsymbol{h}_{i,k}^{aff,proj}$ is the corresponding coefficient matrix/vector for $\mathbb{U}_{i,k}^{aff}$.

Combining (33.a) and (33.b), we can derive (33.c):

$$\begin{cases} \boldsymbol{H}_{i,k}^{aff,proj}\boldsymbol{\Gamma}_{i,k}^{aff} = \boldsymbol{H}_i^{base,proj} \\ \boldsymbol{h}_{i,k}^{aff,proj} = \boldsymbol{h}_i^{base,proj} + \boldsymbol{H}_{i,k}^{aff,proj}\boldsymbol{\gamma}_{i,k}^{aff} \end{cases} \tag{33.c}$$

According to [39], the containment constraint $\mathbb{U}_{i,k}^{aff} \subseteq \mathbb{U}_{i,k}$ can be expressed as follows:

$$\begin{cases} \boldsymbol{\Lambda}_{i,k} \ge 0 \\ \boldsymbol{\Lambda}_{i,k}\boldsymbol{H}_{i,k}^{aff,proj} = \boldsymbol{H}_{i,k}^{proj} \\ \boldsymbol{\Lambda}_{i,k}\boldsymbol{h}_{i,k}^{aff,proj} \le \boldsymbol{h}_{i,k}^{proj} \end{cases} \tag{33.d}$$

where $\boldsymbol{\Lambda}_{i,k}$ is the auxiliary variable matrix.

Combining (33.c) and (33.d), we can derive (23.a)-(23.c). ■

*D. Derivation of Disaggregation Results*

Let $\boldsymbol{P}_i^{agg} \in \mathbb{P}_i^{agg}$ denote an arbitrary point in $\mathbb{P}_i^{agg}$. Recall that $\mathbb{P}_i^{agg} = \sum_k(\boldsymbol{\Gamma}_{i,k}^{aff})\,\mathbb{U}_i^{base} + \sum_k(\boldsymbol{\gamma}_{i,k}^{aff}) \subseteq \mathbb{U}_i^{agg}$ as shown in (14.c). Thus, there exists a point $\boldsymbol{P}_i^{a,base} \in \mathbb{U}_i^{base}$ such that the given point $\boldsymbol{P}_i^{a,base}$ can be expressed as (34.a):

$$\boldsymbol{P}_i^{agg} = \left( \sum_k (\boldsymbol{\Gamma}_{i,k}^{aff}) \right)\boldsymbol{P}_i^{a,base} + \left( \sum_k (\boldsymbol{\gamma}_{i,k}^{aff}) \right) \tag{34.a}$$

Also, $\boldsymbol{\Gamma}_{i,k}^{aff}\mathbb{U}_i^{base} + \boldsymbol{\gamma}_{i,k}^{aff} \subseteq \mathbb{U}_{i,k}^{AC}$ holds, implying that $\boldsymbol{\Gamma}_{i,k}^{aff}\boldsymbol{P}_i^{a,base} + \boldsymbol{\gamma}_{i,k}^{aff} \subseteq \mathbb{U}_{i,k}^{AC}$ holds for $\forall \boldsymbol{P}_i^{a,base} \in \mathbb{U}_i^{base}$. Thus, $\boldsymbol{P}_i^{agg}$ can be disaggregated into a collection of individually feasible points $\boldsymbol{P}_{i,k}^{a}$ , given by (34.b):

$$\boldsymbol{P}_i^{agg} = \sum_k (\boldsymbol{P}_{i,k}^{a}), \boldsymbol{P}_{i,k}^{a} = \boldsymbol{\Gamma}_{i,k}^{aff}\boldsymbol{P}_i^{a,base} + \boldsymbol{\gamma}_{i,k}^{aff} \tag{34.b}$$

Since the parameter value of $\boldsymbol{\Gamma}_{i,k}^{aff}$ is derived by maximizing the $\det(\boldsymbol{\Gamma}_{i,k}^{aff})$ as shown in (16.a), the optimal $\boldsymbol{\Gamma}_{i,k}^{aff}$ must be positive definite. Thus, the inverse transformation $\left(\sum_k(\boldsymbol{\Gamma}_{i,k}^{aff})\right)^{-1}$ exists. Then, combining (34.a) and (34.b), the disaggregated power profiles $\boldsymbol{P}_{i,k}^{a}$ can be expressed as explicit functions of the aggregate power profile $\boldsymbol{P}_i^{agg}$ as follows:

$$\boldsymbol{P}_{i,k}^{a} = \boldsymbol{\gamma}_{i,k}^{aff} + \boldsymbol{\Gamma}_{i,k}^{aff}\left( \sum_k (\boldsymbol{\Gamma}_{i,k}^{aff}) \right)^{-1}\left( \boldsymbol{P}_i^{agg} - \sum_k (\boldsymbol{\gamma}_{i,k}^{aff}) \right) \tag{34.c}$$

Since $\boldsymbol{\Gamma}_{i,k}^{aff}$ is a diagonal matrix, we can further recast (34.c) as a period-decoupled form for real-time scheduling, as shown below:

$$P_{i,k,t}^{a} = \gamma_{i,k,t}^{aff} + \Gamma_{i,k,t}^{aff}\left( \sum_k (\Gamma_{i,k,t}^{aff}) \right)^{-1}\left( P_{i,t}^{agg} - \sum_k (\gamma_{i,k,t}^{aff}) \right) \tag{34.d}$$

## References

[1] J. Zhu *et al.*, "Stochastic Energy Management of Active Distribution Network Based on Improved Approximate Dynamic Programming," *IEEE Trans. Smart Grid*, vol. 13, no. 1, pp. 406–416, Jan. 2022, doi: 10.1109/TSG.2021.3111029.

[2] J. Liu *et al.*, "Continuous-Time Aggregation of Massive Flexible HVAC Loads Considering Uncertainty for Reserve Provision in Power System Dispatch," *IEEE Trans. Smart Grid*, pp. 1–1, 2024, doi: 10.1109/TSG.2024.3398627.

[3] Z. Zhang, H. Hui, and Y. Song, "Response Capacity Allocation of Air Conditioners for Peak-Valley Regulation Considering Interaction with Surrounding Microclimate," *IEEE Trans. Smart Grid*, pp. 1–1, 2024, doi: 10.1109/TSG.2024.3482361.

[4] R. Liu, H. Hui, X. Chen, and Y. Song, "Distributed Frequency Control of Heterogeneous Energy Storage Systems Considering Short-Term Ability and Long-Term Flexibility," *IEEE Trans. Smart Grid*, vol. 15, no. 6, pp. 5693–5705, Nov. 2024, doi: 10.1109/TSG.2024.3451614.

[5] L. Chen and H. Hui, "Model Predictive Control-Based Active/Reactive Power Regulation of Inverter Air Conditioners for Improving Voltage Quality of Distribution Systems," *IEEE Trans. Ind. Inf.*, pp. 1–10, 2024, doi: 10.1109/TII.2024.3468475.

[6] M. Song, W. Sun, M. Shahidehpour, M. Yan, and C. Gao, "Multi-Time Scale Coordinated Control and Scheduling of Inverter-Based TCLs With Variable Wind Generation," *IEEE Trans. Sustain. Energy*, vol. 12, no. 1, pp. 46–57, Jan. 2021, doi: 10.1109/TSTE.2020.2971271.

[7] S. Su, Z. Li, X. Jin, K. Yamashita, M. Xia, and Q. Chen, "Energy management for active distribution network incorporating office buildings based on chance-constrained programming," *INT J ELEC POWER*, vol. 134, p. 107360, Jan. 2022, doi: 10.1016/j.ijepes.2021.107360.

[8] W. Aslam, P. Andrianesis, and M. Caramanis, "Optimal HVAC Energy and Regulation Reserve Scheduling in Power Markets," *IEEE Trans. Sustain. Energy*, vol. 15, no. 1, pp. 201–214, Jan. 2024, doi: 10.1109/TSTE.2023.3279060.

[9] Y. Ma, W. Xu, H. Yang, and D. Zhang, "Two-stage stochastic robust optimization model of microgrid day-ahead dispatching considering controllable air conditioning load," *INT J ELEC POWER*, vol. 141, p. 108174, Oct. 2022, doi: 10.1016/j.ijepes.2022.108174.

[10] L. Le, J. Fang, X. Ai, S. Cui, and J. Wen, "Aggregation and Scheduling of Multi-Chiller HVAC Systems in Continuous-Time Stochastic Unit Commitment for Flexibility Enhancement," *IEEE Trans. Smart Grid*, vol. 14, no. 4, pp. 2774–2785, 2022, doi: 10.1109/TSG.2022.3227390.

[11] H. Saberi, C. Zhang, and Z. Y. Dong, "Data-Driven Distributionally Robust Hierarchical Coordination for Home Energy Management," *IEEE Trans. Smart Grid*, vol. 12, no. 5, pp. 4090–4101, Sep. 2021, doi: 10.1109/TSG.2021.3088433.

[12] Y. Zhang, S. Shen, and J. Mathieu, "Distributionally Robust Chance-Constrained Optimal Power Flow with Uncertain Renewables and Uncertain Reserves Provided by Loads," *IEEE Trans. Power Syst.*, pp. 1–1, 2016, doi: 10.1109/TPWRS.2016.2572104.

[13] Z. Guo, W. Wei, L. Chen, M. Shahidehpour, and S. Mei, "Distribution System Operation With Renewables and Energy Storage: A Linear Programming Based Multistage Robust Feasibility Approach," *IEEE Trans. Power Syst.*, vol. 37, no. 1, pp. 738–749, Jan. 2022, doi: 10.1109/TPWRS.2021.3095281.

[14] H. Qiu, W. Gu, C. Ning, X. Lu, P. Liu, and Z. Wu, "Multistage Mixed-Integer Robust Optimization for Power Grid Scheduling: An Efficient Reformulation Algorithm," *IEEE Trans. Sustain. Energy*, vol. 14, no. 1, pp. 254–271, Jan. 2023, doi: 10.1109/TSTE.2022.3210214.

[15] Y. Zhou, Q. Zhai, and L. Wu, "Multistage Transmission-Constrained Unit Commitment With Renewable Energy and Energy Storage: Implicit and Explicit Decision Methods," *IEEE Trans. Sustain. Energy*, vol. 12, no. 2, pp. 1032–1043, Apr. 2021, doi: 10.1109/TSTE.2020.3031054.

[16] Z. Li, L. Wu, Y. Xu, S. Moazeni, and Z. Tang, "Multi-Stage Real-Time Operation of a Multi-Energy Microgrid With Electrical and Thermal Energy Storage Assets: A Data-Driven MPC-ADP Approach," *IEEE Trans. Smart Grid*, vol. 13, no. 1, pp. 213–226, Jan. 2022, doi: 10.1109/TSG.2021.3119972.

[17] R. Lu, T. Ding, B. Qin, J. Ma, X. Fang, and Z. Dong, "Multi-Stage Stochastic Programming to Joint Economic Dispatch for Energy and Reserve With Uncertain Renewable Energy," *IEEE Trans. Sustain. Energy*, vol. 11, no. 3, pp. 1140–1151, Jul. 2020, doi: 10.1109/TSTE.2019.2918269.

[18] F. Nematkhah, S. Bahrami, F. Aminifar, and J. P. S. Catalao, “Exploiting the Potentials of HVAC Systems in Transactive Energy Markets,” *IEEE Trans. Smart Grid*, vol. 12, no. 5, pp. 4039–4048, Sep. 2021, doi: 10.1109/TSG.2021.3078655.
[19] A. Das, D. Wu, B. A. Bhatti, and M. Kamaludeen, “Approximate Dynamic Programming with Enhanced Off-policy Learning for Coordinating Distributed Energy Resources,” *IEEE Trans. Sustain. Energy*, pp. 1–13, 2024, doi: 10.1109/TSTE.2024.3361674.
[20] X. Xue *et al.*, “Real-Time Schedule of Microgrid for Maximizing Battery Energy Storage Utilization,” *IEEE Trans. Sustain. Energy*, vol. 13, no. 3, pp. 1356–1369, Jul. 2022, doi: 10.1109/TSTE.2022.3153609.
[21] W. Zhang *et al.*, “Proactive Security-Constrained Unit Commitment Against Typhoon Disasters: An Approximate Dynamic Programming Approach,” *IEEE Trans. Ind. Inf.*, vol. 19, no. 5, pp. 7076–7087, May 2023, doi: 10.1109/TII.2022.3208574.
[22] B. Zheng, W. Wei, Y. Xu, and Y. Chen, “Capacity Aggregation and Online Control of Clustered Energy Storage Units,” *IEEE Trans. Sustain. Energy*, pp. 1–15, 2024, doi: 10.1109/TSTE.2024.3355991.
[23] E. Öztürk, T. Faulwasser, K. Worthmann, M. PreißInger, and K. Rheinberger, “Alleviating the Curse of Dimensionality in Minkowski Sum Approximations of Storage Flexibility,” *IEEE Trans. Smart Grid*, vol. 15, no. 6, pp. 5733–5743, Nov. 2024, doi: 10.1109/TSG.2024.3420156.
[24] M. Zhang, Y. Xu, and Z. Yi, “Two-stage Carbon-Oriented Scheduling of an Active Distribution Network with Thermostatically Controlled Load Aggregators,” *IEEE Trans. Sustain. Energy*, pp. 1–13, 2024, doi: 10.1109/TSTE.2024.3351720.
[25] M. Zhang, Y. Xu, X. Shi, and Q. Guo, “A Fast Polytope-based Approach for Aggregating Large-Scale Electric Vehicles in the Joint Market Under Uncertainty,” *IEEE Trans. Smart Grid*, pp. 1–1, 2023, doi: 10.1109/TSG.2023.3274198.
[26] J. Han, Y. Fang, E. Du, P. Yong, N. Zhang, and N. Liu, “Eliminating Distribution Network Congestion Based on Spatial-Temporal Migration of Multiple Base Stations,” *IEEE Trans. Smart Grid*, vol. 15, no. 6, pp. 5638–5652, Nov. 2024, doi: 10.1109/TSG.2024.3418976.
[27] M. Yu, S. H. Hong, Y. Ding, and X. Ye, “An Incentive-Based Demand Response (DR) Model Considering Composited DR Resources,” *IEEE Trans. Ind. Electron.*, vol. 66, no. 2, pp. 1488–1498, Feb. 2019, doi: 10.1109/TIE.2018.2826454.
[28] J. Yang, N. Zhang, C. Kang, and Q. Xia, “A State-Independent Linear Power Flow Model With Accurate Estimation of Voltage Magnitude,” *IEEE Trans. Power Syst.*, vol. 32, no. 5, pp. 3607–3617, Sep. 2017, doi: 10.1109/TPWRS.2016.2638923.
[29] H. Shuai, J. Fang, X. Ai, Y. Tang, J. Wen, and H. He, “Stochastic Optimization of Economic Dispatch for Microgrid Based on Approximate Dynamic Programming,” *IEEE Trans. Smart Grid*, vol. 10, no. 3, pp. 2440–2452, May 2019, doi: 10.1109/TSG.2018.2798039.
[30] J. Nascimento and W. B. Powell, “An Optimal Approximate Dynamic Programming Algorithm for Concave, Scalar Storage Problems With Vector-Valued Controls,” *IEEE Trans. Automat. Contr.*, vol. 58, no. 12, pp. 2995–3010, Dec. 2013, doi: 10.1109/TAC.2013.2272973.
[31] W. B. Powell, *Approximate Dynamic Programming: Solving the Curses of Dimensionality*, 2nd ed., 2 vols. Hoboken, NJ, 2011.
[32] L. Zhao, W. Zhang, H. Hao, and K. Kalsi, “A Geometric Approach to Aggregate Flexibility Modeling of Thermostatically Controlled Loads,” *IEEE Trans. Power Syst.*, vol. 32, no. 6, pp. 4721–4731, Nov. 2017, doi: 10.1109/TPWRS.2017.2674699.
[33] J. Jian, M. Zhang, Y. Xu, W. Tang, and S. He, “An Analytical Polytope Approximation Aggregation of Electric Vehicles Considering Uncertainty for the Day-Ahead Distribution Network Dispatching,” *IEEE Trans. Sustain. Energy*, pp. 1–12, 2023, doi: 10.1109/TSTE.2023.3275566.
[34] C. Keerthisinghe, G. Verbic, and A. C. Chapman, “A Fast Technique for Smart Home Management: ADP With Temporal Difference Learning,” *IEEE Trans. Smart Grid*, vol. 9, no. 4, pp. 3291–3303, Jul. 2018, doi: 10.1109/TSG.2016.2629470.
[35] J. Nascimento and W. B. Powell, “An Optimal Approximate Dynamic Programming Algorithm for Concave, Scalar Storage Problems With Vector-Valued Controls,” *IEEE TRANSACTIONS ON AUTOMATIC CONTROL*, vol. 58, no. 12, 2013.
[36] “System Parameter of 8-bus and 123-bus System.” Dec. 28, 2024. [Online]. Available: https://zenodo.org/records/10720090
[37] X. Xue *et al.*, “A Fully Distributed ADP Algorithm for Real-time Economic Dispatch of Microgrid,” *IEEE Trans. Smart Grid*, pp. 1–1, 2023, doi: 10.1109/TSG.2023.3273418.
[38] R. Kamyar and M. M. Peet, “Optimal Thermostat Programming for Time-of-Use and Demand Charges With Thermal Energy Storage and Optimal Pricing for Regulated Utilities,” *IEEE Trans. Power Syst.*, vol. 32, no. 4, pp. 2714–2723, Jul. 2017, doi: 10.1109/TPWRS.2016.2618374.
[39] B. Liu and J. H. Braslavsky, “Robust Dynamic Operating Envelopes for DER Integration in Unbalanced Distribution Networks,” *IEEE Trans. Power Syst.*, pp. 1–15, 2023, doi: 10.1109/TPWRS.2023.3308104.

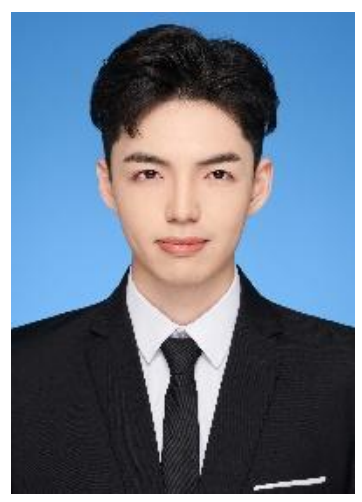

**Jingguan Liu** (Student Member, IEEE) received the B.E. degree in 2022 in electrical engineering from the Huazhong University of Science and Technology, Wuhan, China, where he is currently working toward the Ph.D. degree in electrical engineering. His current research interests include aggregation of flexible loads, flexible operation of power systems, and continuous-time scheduling.

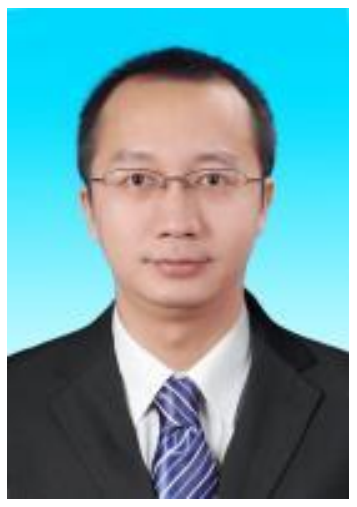

**Xiaomeng Ai** (Member, IEEE) received the B.S. degree in mathematics and applied mathematics and the Ph.D. degree in electrical engineering from the Huazhong University of Science and Technology (HUST), Wuhan, China, in 2008 and 2014 respectively. He is currently a Professor with the School of Electrical and Electronics Engineering, HUST. His research interests include robust optimization theory in power system, renewable energy integration, and integrated energy market.

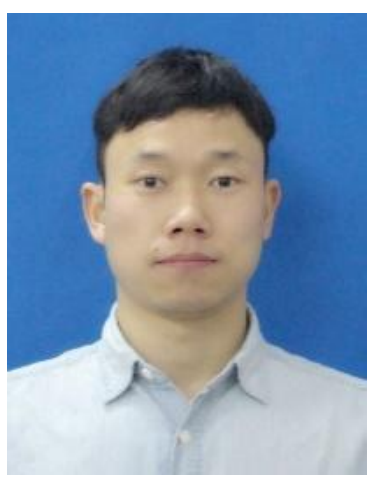

**Shichang Cui** (Member, IEEE) received the B.E. degree in Automation and the Ph.D. degree in control science and engineering from Huazhong University of Science and Technology, Wuhan, China, in 2016 and 2021, respectively. He currently works as an associate research fellow in the State Key Laboratory of Advanced Electromagnetic Technology, Huazhong University of Science and Technology, Wuhan, 430074, China. From Sep. 2019 to Sep. 2020, he was visiting the department of Mechanical Engineering, University of Victoria, Canada, supported by the CSC Joint Doctoral Program. His current research interests include stochastic optimization, distributed optimization, game theory, and energy management for smart grids.

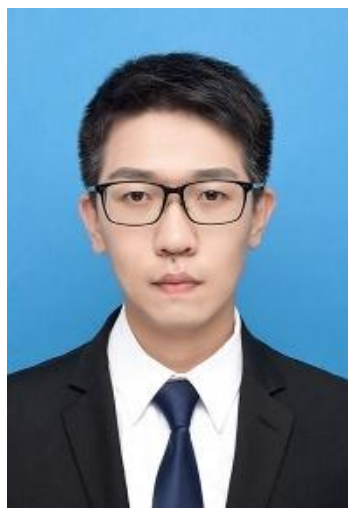

**Xizhen Xue** (Member, IEEE) received the B.E. and Ph.D. degrees in electrical engineering from the Huazhong University of Science and Technology, Wuhan, China, in 2019 and 2024, respectively. From 2023 to 2024, he was a Visiting Research Scholar with the Department of Electrical and Computer Engineering, Clarkson University, Potsdam, NY, USA. He is currently a Research Fellow with the School of Electrical and Electronic Engineering, Nanyang Technological University, Singapore. His current research interests include approximate dynamic programming, distributed optimization algorithm, energy storage scheduling and planning, and integrated energy systems.

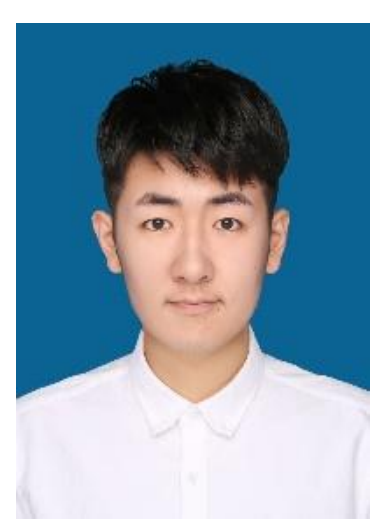

**Shengshi Wang** (Student Member, IEEE) received the B.E. degree in electrical engineering from Chongqing University, Chongqing, China, in June 2020. He is currently pursuing the Ph.D. degree in electrical engineering with the Huazhong University of Science and Technology, Wuhan, China. From September 2023 to November 2024, he was a visiting student at Cardiff University, Cardiff, United Kingdom. His research interests lie in decision theory—particularly robust optimization with decision-dependent uncertainties, approximate dynamic programming, and deep reinforcement learning, with a focus on their applications in the energy systems.

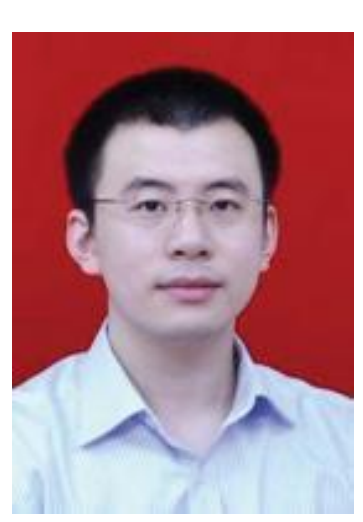

**Jiakun Fang** (Senior Member, IEEE) received the B.E. and Ph.D. degrees from the Huazhong University of Science and Technology (HUST), Wuhan, China, in 2007 and 2012, respectively. From 2012 to 2019, he was with the Department of Energy Technology, Aalborg University, Aalborg, Denmark. He is currently a Professor with the School of Electrical and Electronics Engineering, Huazhong University of Science and Technology. His research interests include the optimal integration of the power and gas systems, and the storage across multiple energy carriers.

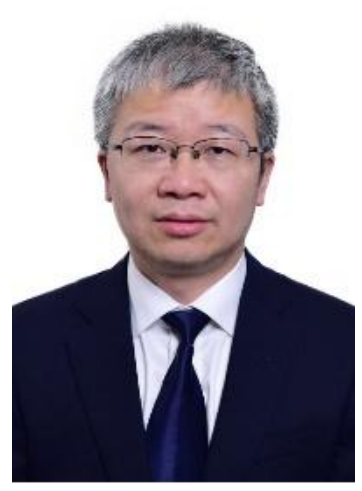

**Jinyu Wen** (Member, IEEE) received the B.E. and Ph.D. degrees in electrical engineering from the Huazhong University of Science and Technology, Wuhan, China, in 1992 and 1998, respectively. He was a Visiting Student from 1996 to 1997, and a Research Fellow from 2002 to 2003, with the University of Liverpool, Liverpool, U.K., and a Senior Visiting Researcher with the University of Texas at Arlington, Arlington, TX, USA, in 2010. From 1998 to 2002, he was a Director Engineer with XJ Electric Company Ltd., China. In 2003, he joined HUST, where he is currently a Professor with the School of Electrical and Electronics Engineering. His current research interests include renewable energy integration, energy storage, multiterminal HVDC, and power system operation and control.

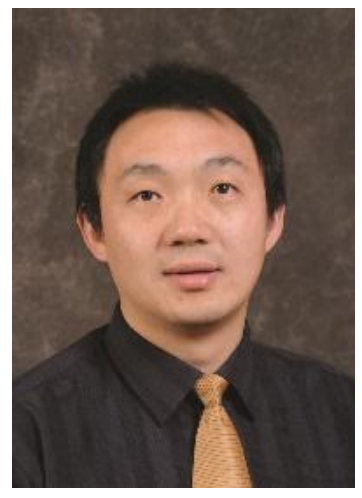

**Yang Shi** (Fellow, IEEE) received the B.Sc. and Ph.D. degrees in mechanical engineering and automatic control from Northwestern Polytechnical University, Xi'an, China, in 1994 and 1998, respectively, and the Ph.D. degree in electrical and computer engineering from the University of Alberta, Edmonton, AB, Canada, in 2005. He is currently a Professor with the Department of Mechanical Engineering, University of Victoria, Victoria, BC, Canada. His research interests include networked and distributed systems, model predictive control, cyber–physical systems, robotics, autonomous systems, and energy system applications. Dr. Shi was the recipient of numerous awards, including the REACH Award for Graduate Student Supervision (2023), the Humboldt Research Fellowship (2018), and the IEEE Dr.-Ing. Eugene Mittelmann Achievement Award (2023). He is an Editor-in-Chief for IEEE TRANSACTIONS ON INDUSTRIAL ELECTRONICS and IEEE CANADIAN JOURNAL OF ELECTRICAL AND COMPUTER ENGINEERING. He is a Fellow of ASME, CSME, EIC, and the Canadian Academy of Engineering, and is a registered Professional Engineer in British Columbia.